\documentstyle[sprocl,psfigure,rotating]{article}

\def\Journal#1#2#3#4{{#1} {\bf #2}, #3 (#4)}

\def\NIM{\em Nucl. Instrum. Methods}
\def\NIMA{{\em Nucl. Instrum. Methods} A}
\def\NPB{{\em Nucl. Phys.} B}
\def\PLB{{\em Phys. Lett.}  B}
\def\PRL{\em Phys. Rev. Lett.}
\def\PRD{{\em Phys. Rev.} D}
\def\ZPC{{\em Z. Phys.} C}

\def\be{\begin{equation}}
\def\ee{\end{equation}}
\def\bea{\begin{eqnarray}}
\def\eea{\end{eqnarray}}

\begin{document}

\title{THE SPIN STRUCTURE OF THE NUCLEON}

\author{MICHEL C. VETTERLI}

\address{TRIUMF, 4004 Wesbrook Mall, Vancouver, B.C. V6T~2A3, Canada}

\maketitle 

\abstracts{This paper gives a pedagogical introduction to our
knowledge of the spin structure of the nucleon.  In particular,
polarised deep inelastic lepton scattering is presented as a tool to
study how the nucleon's constituents combine to generate its spin.  The
importance of semi-inclusive measurements is discussed and a window on
future experiments in the field is given.}

\section{Introduction}

This paper gives an overview of the spin structure of the nucleon with
emphasis on a pedagogical presentation of the tools used to study this
fundamental property of protons and neutrons.  While many of the data
currently available are shown, this paper is not intended to be an
exhaustive summary of the field.  Rather, the data are used to
illustrate the concepts presented.  The most recent progress in the
field can be found in the proceedings of the annual conference on deep
inelastic scattering and QCD~\cite{dis96,dis97}.

The last ten years have seen a wealth of measurements of the
polarisation asymmetry in the cross-section for deep inelastic lepton
scattering (DIS).  This has been made possible by impressive advances
in polarised target and beam technologies. Precise data now exist on
{\em inclusive} polarised DIS from experiments at SLAC, CERN, and
DESY.  These data can be interpreted as showing that surprisingly, only
about 30\% of the nucleon's spin comes from the spins of the quarks.
To make further progress on this problem, {\em semi-inclusive}
measurements are needed where hadrons in coincidence with the
scattered lepton are detected.  These data allow us to determine the
flavour of the quark that was struck and isolate the contribution to
the nucleon's spin by valence and sea quarks.  Furthermore,
semi-inclusive data on charm production should give information on the
contribution of gluons to the nucleon's spin.  Current and future
experiments are concentrating on semi-inclusive measurements.

After a short discussion of magnetic moments, the formalism of
polarised deep inelastic scattering is presented in
Sec.~\ref{subsec:formalism}-\ref{subsec:practice}.  The data on the
polarised structure function $g_1(x)$ are then shown in
Sec.~\ref{sec:g1data} which is followed by the interpretation of these
data in terms of sum rules in Sec.~\ref{sec:sumrules}.
Section~\ref{sec:content} shows how spin dependent structure functions
can be used to determine the contribution of the quark spins to the
spin of the nucleon.  The $Q^2$ evolution of $g_1(x)$ is discussed in
Sec.~\ref{sec:q2}, while other spin structure functions are presented
in Sec.~\ref{sec:others}.  A description of the experiments which have
addressed the nucleon spin problem over the last decade is given in
Sec.~\ref{sec:exp}.  Semi-inclusive measurements, which are the
current focus of the field are discussed in section~\ref{sec:semiinc}.
Finally, the future directions of the field are discussed in
section~\ref{sec:future}.

\section{The Spin of Three-Body States}

As a preamble to the discussion on polarised deep inelastic
scattering, two simple examples of how three spin 1/2 particles
combine to form a spin 1/2 system are presented: the ground-state of
$^3$He and $^3$H, and baryon magnetic moments.

\subsection{The Nuclear 3-Body Ground State}\label{subsec:3he}

The wave functions of $^3$He and $^3$H are among the most studied
topics in physics.  $^3$He is made up of two protons and one neutron,
each of which has a spin of 1/2. $^3$H has one proton and two
neutrons.  The total wavefunction also has a spin of 1/2 leading to
the following possibilities for the combination of the three nucleons:

\begin{center}
\begin{tabular}{lll}
L=0, S= 1/2      & $\Rightarrow$ & S-State ($>$90\%) \\
L=1, S= 1/2, 3/2 & $\Rightarrow$ & P-State ($<$1\%)  \\
L=2, S= 3/2      & $\Rightarrow$ & D-State (8-9\%).
\end{tabular}
\end{center}

\noindent
L is the orbital angular momentum of the 3-body system, while S is the
total spin.  Since the ground-state is overwhelmingly L=0, the two
protons are anti-aligned due to the Pauli exclusion principle, and any
spin effects observed in $^3$He should come from the unpaired neutron.
For example, consider the magnetic moment given by:

\begin{center}
\begin{tabular}{ll}
$\vec{\mu_s}= g_s \cdot \vec{s}$\ \ ;\ \ & $g_s$(p)= 5.5855, \\
                                         & $g_s$(n)= --3.8327.
\end{tabular}
\end{center}

\noindent
We would have $\rm \mu_s(^3He)=\mu_s(n)$ and $\rm \mu_s(^3H)= \mu_s(p)$
if the wave functions were purely S-state.  This is approximately true
in reality:

\begin{center}
\begin{tabular}{lllr}
$\rm \mu_s(^3He)= -2.2174$ & $\approx$ & $\mu_s$(n)= --1.9130 \\
$\rm \mu_s(^3H)= 2.9788$ & $\approx$ & $\mu_s$(p)= 2.7927.
\end{tabular}
\end{center}

\noindent
A large part of the difference can be ascribed to meson exchange
currents.  Since these have opposite sign in the two nuclei, they
should cancel in the average of the magnetic moments:

\begin{center}
$\rm \frac{1}{2}[\mu_s(^3H) + \mu_s(^3He)]$=  0.4257 \\
$\rm \frac{1}{2}[\mu_s(p) + \mu_s(n)]$=  0.4399.
\end{center}

\noindent
The remaining difference ($\approx$3.2\%) is due to the other components
of the ground-state wavefunction (P- and D-states).

As an experimental aside, the fact that polarised $^3$He looks very
much like a polarised neutron (at least as far as spin effects are
concerned) makes it an ideal polarised neutron target.  This
has been exploited both at SLAC (E142/E154) and at DESY (HERMES).

\begin{figure}
\hspace{1.0cm}\psfig{figure=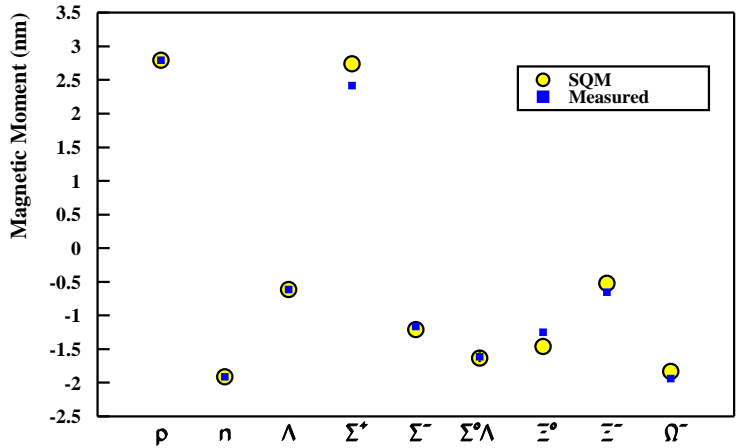,width=4.0in}
\caption{Baryon magnetic moments.  The proton, neutron,
and $\Lambda$ are used as input parameters, while the other magnetic
moments are predictions.  The agreement of the data with the simple
quark model (SQM) is very good.
\label{fig:magmom}}
\end{figure}

\subsection{Baryon Magnetic Moments}\label{subsec:baryon}

Since the simple model of three spin 1/2 particles combining to form a
spin 1/2 state works so well in the nuclear case, it is natural to
apply it to the nucleon.  However, in this case we do not know the
magnetic moments of the constituents, the quarks, since they cannot be
isolated.  Nevertheless, we can test the simple model by
looking at the ratio of $\mu_s$(p) and $\mu_s$(n) where the quark
magnetic moments cancel.  Experimentally, $\mu_s$(p)/$\mu_s$(n)=
--1.46, while the simple quark parton model predicts --1.50.
Furthermore, we can use the proton, the neutron, and the $\Lambda$ to
set the values of $\mu_s$(u), $\mu_s$(d), $\mu_s$(s), and predict the
magnetic moments of the other baryons.  The
results are shown in Fig.~\ref{fig:magmom}.  The agreement between the
predictions and the measurements is very good indicating that the
simple model works well.  However, while the magnetic moments are
consistent with the model presented here, they depend only on the
static properties of the nucleon wavefunction (more precisely at
$Q^2=0$) and the rest of this paper will discuss how this picture is
too simple.

\section{Inclusive Polarised Deep Inelastic Scattering}\label{sec:PDIS}
\subsection{Polarised DIS Formalism}\label{subsec:formalism}

\begin{figure}
\hspace{1.0cm}\psfig{figure=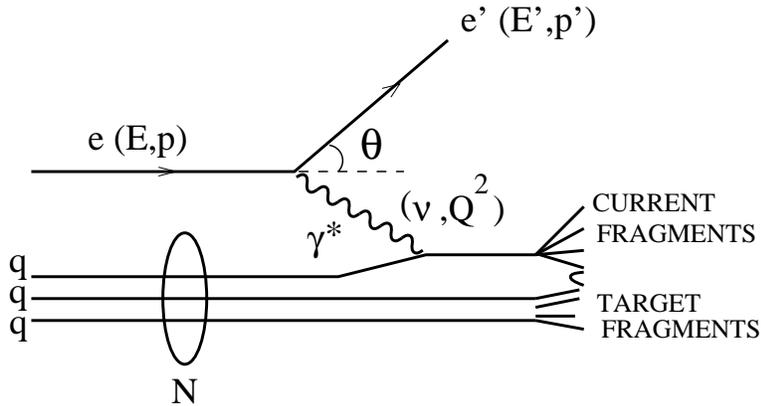,width=4.0in}
\caption{Diagram of the Deep Inelastic Scattering process.  The incoming
lepton emits a virtual photon which is absorbed by one of the quarks in
the nucleon.
\label{fig:DIS}}
\end{figure}

The main process we will be concerned with in these lectures, deep
inelastic scattering, is depicted in Fig.~\ref{fig:DIS}.  An incoming
lepton (shown here as a positron or an electron) emits a virtual
photon which is absorbed by a quark in the nucleon.  The nucleon is
broken up and the struck quark and the target remnant fragment into
hadrons in the final state.  Only the lepton is detected in inclusive
measurements while detection of one or more hadrons in the final state
(semi-inclusive data) adds important information on the scattering
process.  W$^\pm$ and Z$^0$ exchanges are not important at the
energies of the current spin experiments.  The kinematic variables
relevant for this process are listed in Table~\ref{tab:DIS}.  The
formalism for Deep Inelastic Scattering (DIS) is developped in many
texts on particle
physics~\cite{perkins,halzen,close,griffiths,roberts}.

\begin{table}
\caption{Kinematic variables relevant to Deep Inelastic Scattering}

\ \

\begin{tabular}{|ll|}
\hline
$(\vec{k},E)$        & Momentum and energy of the incoming lepton \\
$(\vec{k}',E')$      & Momentum and energy of the outgoing lepton \\
$\nu= E-E'$          & Energy transfer in the reaction \\
$q= k-k'$            & Four momentum transfer \\
$Q^2=-q^2$           & = 4$EE'$ sin$^2(\theta$/2); \, $\theta$: e$^\pm$
scattering angle \\
$x= Q^2/2M_p\nu$     & Fraction of the nucleon momentum carried by \\ 
                     & the struck quark \\
$y= \nu/E$           & Fraction of the incident energy transferred \\
$W^2= M_p^{2}+2M_p\nu-Q^2$ & Square of the invariant mass of the \\
                     & photon-nucleon system \\
$z= E_h/\nu$         & Fraction of the energy transferred carried by \\
                     & a particular hadron $h$ \\
$x_F= 2p_\parallel/W$ & Scaled longitudinal momentum of a hadron \\
\hline
\end{tabular}
\label{tab:DIS}
\end{table}

The DIS cross-section can be written as follows:
\begin{equation}
\frac{d^2\sigma}{d\Omega~dE'}= \, \frac{\alpha^2}{2M_pQ^4} \, \frac{E'}{E}
\, L_{\mu\nu}~W^{\mu\nu}
\end{equation}
where $\alpha$ is the fine structure constant and $M_p$ is 
the mass of the proton.  Notice that
the cross-section falls off very quickly as a function of momentum
transfer Q.  $L_{\mu\nu}$ is a tensor which describes the emission of
the virtual photon by the lepton; it can be calculated exactly in
QED. $W^{\mu\nu}$ is a tensor describing the absorption of the virtual
photon by the target; it contains all of the information related to
the structure of the target.  Symmetry considerations and conservation
laws determine the form of $W^{\mu\nu}$:
\begin{eqnarray}
\label{eqn:wmunu}
W^{\mu\nu} & = & -g^{\mu\nu} \cdot {\bf F_1}
\mbox{} \ \ + \ \ p^\mu p^\nu / \nu \cdot {\bf F_2} \\
 & & \mbox{} + i/\nu~\epsilon^{\mu\nu\lambda\sigma}~q^{\lambda}
s^{\sigma} \cdot {\bf g_1}
\mbox{} \ \ + \ \ i/\nu^2~\epsilon^{\mu\nu\lambda\sigma}~q^{\lambda} 
(p.q~s^{\sigma}~-~s.q~p^{\sigma}) \cdot {\bf g_2}. \nonumber
\end{eqnarray}

\noindent
$F_1$ and $F_2$ are {\em unpolarised} structure functions, while $g_1$
and $g_2$ are {\em polarised} structure functions which
contribute to the cross-section only if both the target and the beam
are polarised.  In general, the structure functions depend on
$\nu$ and $Q^2$.  However, if the scattering occurs from pointlike
constituents, they depend only on one variable $x=~Q^2/2M_p\nu$.  This
is the phenomenon of scaling which was the first experimental
confirmation that the nucleon is composed of smaller particles.  In
the infinite momentum frame, $x$ is shown to be the fraction of the
nucleon's momentum carried by the struck quark.  The structure
functions can be interpreted in the quark-parton model in a
straightforward way:
\begin{equation}
F_1(x)=~ \frac{1}{2}~\sum_{f}e_f^2~(q_f^+(x)+q_f^-(x))~=
\frac{1}{2}~\sum_{f}e_f^2~q_f(x)
\label{eqn:f1}
\end{equation}
\begin{equation}
g_1(x)=~ \frac{1}{2}~\sum_{f}e_f^2~(q_f^+(x)-q_f^-(x))~=~
\frac{1}{2}~\sum_{f}e_f^2~\Delta q_f(x)
\label{eqn:g1}
\end{equation}

\noindent
where the sum is over quark flavours and $e_f$ is the charge of quark.
$q_f^+$ ($q_f^-$) is the distribution of quarks {\em and} anti-quarks with
their spins in the same (opposite) direction as the spin of the
nucleon.  $F_1(x)$ measures the momentum distribution of the quarks in
the nucleon, while $g_1(x)$ gives the distribution of the spins of the
quarks.  For spin 1/2 quarks, $F_2$ is related to $F_1$ by~\cite{ca69}
\begin{equation}
F_2(x)= 2 x F_1(x).
\end{equation}
Finally, $g_2=0$ in the quark-parton model since it is related to transverse
degrees of freedom which are absent in the simple QPM.

\begin{figure}
\hspace{1.0cm}
\psfig{figure=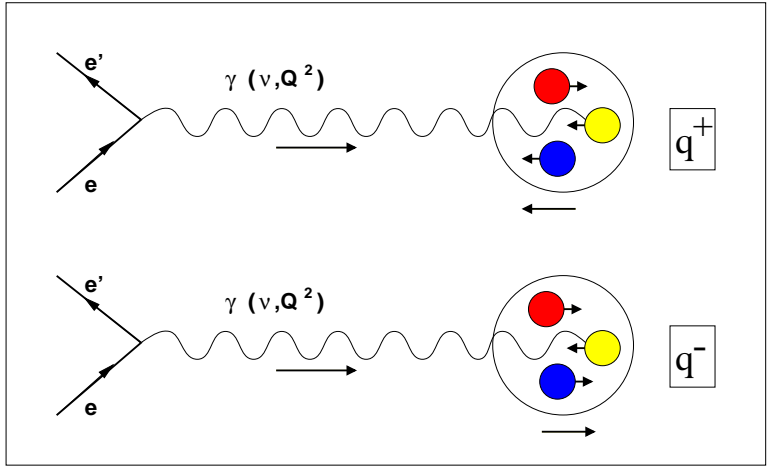,width=4.0in}
\caption{Diagram of Polarised Deep Inelastic Scattering.  The virtual photon
is always absorbed by a quark with opposite spin direction.  The top panel 
shows the process that is sensitive to quarks polarised parallel to the 
nucleon's spin ($q_f^+(x)$), while the bottom panel is for quarks 
with their spin direction anti-parallel to the nucleon ($q_f^-(x)$).
\label{fig:cartoon}}
\end{figure}

An intuitive picture of how $g_1(x)$ is sensitive to spin degrees of
freedom and how to measure it is given in Fig.~\ref{fig:cartoon}.  The
incoming polarised lepton at the left emits a circularly polarised
virtual photon with spin projection pointing to the right in this
picture.  This photon can be absorbed only by a quark with spin
projection in the opposite direction because the final state (a quark)
must have spin 1/2 and can therefore not have spin projection 3/2.  In
the top panel of Fig.~\ref{fig:cartoon} the target polarisation
'points' to the left so that we measure the cross-section for beam and
target polarisations anti-parallel.  This cross-section is labelled
$\sigma^{\uparrow \downarrow}$ and is sensitive to the distribution of
quarks with their spins in the same direction as the spin of the host
nucleon, or $q_f^+(x)$.  In the bottom panel of
Fig.~\ref{fig:cartoon}, the target spin direction has been reversed so
that we measure the cross-section for beam and target polarisations
parallel ($\sigma^{\uparrow \uparrow}$).  However, the elementary
process is not changed.  The photon absorbs on a quark with spin
projection 'pointing' to the left and this process is sensitive to
$q_f^-(x)$.  Since $g_1(x)$ is related to the difference of $q_f^+(x)$
and $q_f^-(x)$ (see Eqn.~\ref{eqn:g1}), it can be deduced by measuring
the polarisation asymmetry:
\begin{equation}
A_\parallel=\frac{\sigma^{\uparrow \downarrow} - \sigma^{\uparrow \uparrow}}
{\sigma^{\uparrow \downarrow} + \sigma^{\uparrow \uparrow}}.
\end{equation}

\begin{figure}
\hspace{2.5cm}\psfig{figure=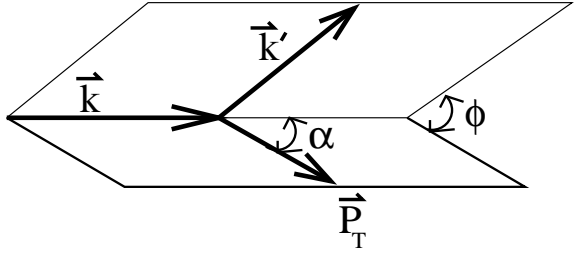,width=3.0in}
\caption{Diagram of the angles relevant to Polarised Deep Inelastic 
Scattering.  See the text for details.
\label{fig:PDIS}}
\end{figure}

Further insight can be gained into how to measure $g_1(x)$ and
$g_2(x)$ by considering Fig.~\ref{fig:PDIS}.  $\vec{k}$ and $\vec{k'}$
are the usual lepton momenta, $\phi$ is the angle between the lepton
scattering plane and the target spin, and $\alpha$ is the angle
between the beam and target spins.  The target polarisation is said to
be longitudinal (transverse) when $\alpha= 0^\circ~{\rm or}~180^\circ
\, (=90^\circ~{\rm or}~270^\circ)$.  The difference in the
cross-sections ($\Delta \sigma$) for the two relative directions of
the beam and target spins can now be written as follows:
\begin{eqnarray}
\frac{d^3\Delta\sigma}{dx dy d\phi}\propto & \cos\alpha
\cdot \{ a~g_1(x) + b~g_2(x) \} \nonumber \\
 & - \cos\phi~\sin\alpha \cdot c \cdot 
\{ \frac{y}{2}~g_1(x) + g_2(x) \}.
\end{eqnarray}

\noindent
Each of the two terms in braces can be isolated by preparing the
target with either longitudinal or transverse target polarisation.
The kinematic constants $a, b,$ and $c$ are of order $\mathcal{O}$(1),
${\mathcal{O}}(1/Q^2)$, and ${\mathcal{O}}(1/Q)$ respectively.
Therefore, $g_1(x)$ dominates for longitudinal target polarisation,
even at moderate $Q^2$.  This statement is strengthened by the fact
that there are good arguments that $g_2(x)$ is small since it is
related to transverse degrees of freedom in the nucleon.  The small
value of $g_2(x)$ has been confirmed by experiment as we will see in
Sec.~\ref{subsec:g2}.  Once $g_1(x)$ has been determined, transverse
target polarisation can be used to maximize the sensitivity to
$g_2(x)$.  However, notice that in this case the asymmetry is reduced
by the factor $c \propto 1/Q$, making this measurement more difficult.

\subsection{What is measured in practice?}\label{subsec:practice}

This section gives the steps required to go from a measurement of the
cross-section asymmetry to the polarised structure functions.  First,
one must realize that the experimentally observed asymmetry ($A_{meas}$)
is reduced because we do not have perfect beam and target
polarisation.  It is related to the true asymmetry ($A_\parallel$) by:
\begin{equation}
A_{meas}= \frac{N ^{\uparrow \downarrow} - N ^{\uparrow \uparrow}} 
{N ^{\uparrow \downarrow} + N ^{\uparrow \uparrow}}= 
(f_D \cdot p_B \cdot p_T) \cdot A_\parallel,
\label{eqn:meas}
\end{equation}

\noindent
where $p_B$ and $p_T$ are respectively the beam and target
polarisations, and $f_D$ is the target dilution factor.  $f_D$ is
defined as the fraction of polarisable nucleons in the target (1 for H
and D gas; $\frac{1}{3}$ for $^3$He gas; $\frac{3}{17}$ for NH$_3$;
and $\frac{10}{74}$ for $\rm C_4H_9OH$).  $f_D$ is further reduced by
extraneous materials in the target such as windows, or vapour
introduced for the optical pumping process.  Beam polarisations vary
from 0.5 to 0.85, while the range of available target polarisations is
0.4-0.9.  The ratio of the measured to the real asymmetry is in the
range 0.07-0.12 (CERN); 0.17-0.35 (SLAC); 0.25-0.50 (DESY) (see
Sec.~\ref{sec:exp}).  Since some of the asymmetries are small to begin
with, the experiments must be very accurate, requiring high
statistics.

The measured asymmetries for longitudinal and transverse target 
polarisation, corrected for the factors in Eqn.~\ref{eqn:meas},
are written in terms of the cross-sections as follows:
\begin{equation}
A_\parallel= \frac{\sigma ^{\uparrow \downarrow} - \sigma ^{\uparrow 
\uparrow}}{\sigma ^{\uparrow \downarrow} + \sigma ^{\uparrow \uparrow}}
\ \ \ \ ;\ \ \ \ 
A_\perp=
\frac{\hat{\sigma} ^{\uparrow \downarrow} - \hat{\sigma} ^{\uparrow 
\uparrow}}{\hat{\sigma} ^{\uparrow \downarrow} + \hat{\sigma} ^{\uparrow 
\uparrow}}, \ \ {\rm where\ } \hat{\sigma}= \int_{}^{} \frac{1}{\cos \phi} 
\frac{d \sigma}{d \phi} d \phi.
\end{equation}

\noindent
While $A_\parallel$ and $A_\perp$ depend on beam and target related
polarised cross-sections, the physically significant asymmetries for
photon absorption on the nucleon level are given by:
\begin{equation}
A_1= \frac{\sigma_{\frac{1}{2}} - \sigma_{\frac{3}{2}}}
{\sigma_{\frac{1}{2}} + \sigma_{\frac{3}{2}}}\ \ \ \ ;\ \ \ \ A_2= 
\frac{\sigma_{TL}}{\frac{1}{2} (\sigma_{\frac{1}{2}} + 
\sigma_{\frac{3}{2}})}= \frac{\sigma_{TL}}{\sigma_T},
\end{equation}

\noindent
where $\sigma_{\frac{1}{2}}$ is the absorption cross-section when the
photon and the nucleon have their spins anti-aligned, and
$\sigma_{\frac{3}{2}}$ the cross-section when they are aligned.
$\sigma_T$ and $\sigma_{TL}$ are the absorption cross-section for
tranversely polarised virtual photons and the longitudinal-transverse
interference respectively.

The measured asymmetries are related to the photon absorption asymmetries
by the following equations:
\begin{equation}
A_\parallel= D \cdot (A_1 + \eta A_2)\ \ \ \ ;\ \ \ \ A_\perp= 
d \cdot (A_2 - \xi A_1).
\end{equation}

\noindent
D, $\eta$, d, and $\xi$ are depolarisation factors given by:
\begin{eqnarray}
D= \frac{y(2-y)}{y^2 + 2(1-y) (1+R)}\ \ \ ;\ \ \ 
\eta= \frac{2 \gamma (1-y)}{2-y}\ \ \ ;\ \ \ 
d= D \sqrt{\frac{2 \epsilon}{1 + \epsilon}} \\
\xi= \eta \frac{1 + \epsilon}{2 \epsilon}\ \ \ ;\ \ \ 
\epsilon= \frac{1-y}{1-y+\frac{y^2}{2}}\ \ \ ;\ \ \ 
\gamma= Q/\nu= 2mx/Q. \nonumber
\end{eqnarray}

\noindent
R is the ratio of longitudinal to transverse
cross-sections~\cite{wi90}.  The depolarisation factors account for
the loss of polarisation from the lepton to the virtual photon and
depend on kinematics.

Finally, the photon asymmetries are related to the structure functions
as follows:
\begin{equation}
A_1= \frac{g_1 - \gamma^2 g_2}{F_1}\ \ \ \ ;\ \ \ \ 
A_2= \frac{\gamma(g_1+g_2)}{F_1}.
\end{equation}

\noindent
If $g_2$ and $A_2$ are ignored, the equation for $g_1$ simplifies
considerably:
\begin{equation}
g_1(x) \approx F_1(x) \cdot A_1(x) \approx F_1(x) \cdot \frac{A_\parallel(x)}
{D}.
\end{equation}

\subsection{Data on the Polarised Structure Function 
$g_1(x)$}\label{sec:g1data}

\begin{figure}
\hspace{0.5cm}\psfig{figure=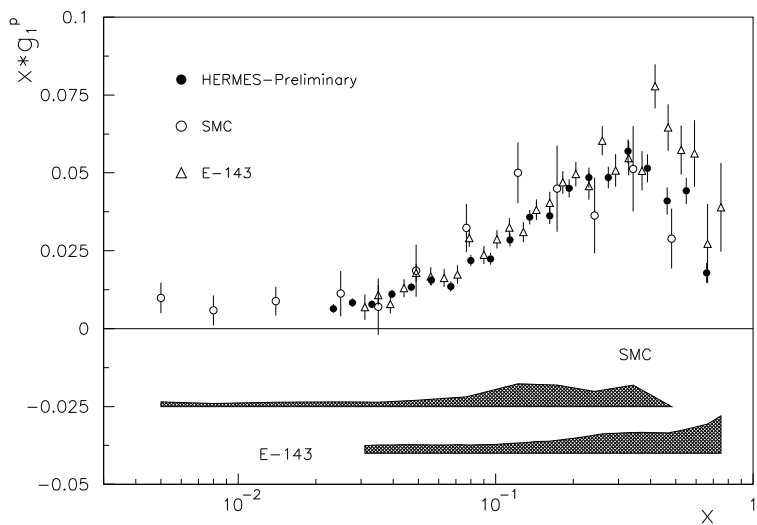,width=10.0cm}
\caption{Data on $x g_1^p(x)$ from SMC~$^{11}$, E143~$^{14}$, 
and preliminary results from HERMES~$^{16}$. The E143 data are at 
$Q^2=$ 3~GeV$^2$, while the SMC and HERMES data are at the measured
$Q^2$ for each $x$ bin.
\label{fig:xg1p}}
\end{figure}

\begin{figure}
\hspace{0.5cm}\psfig{figure=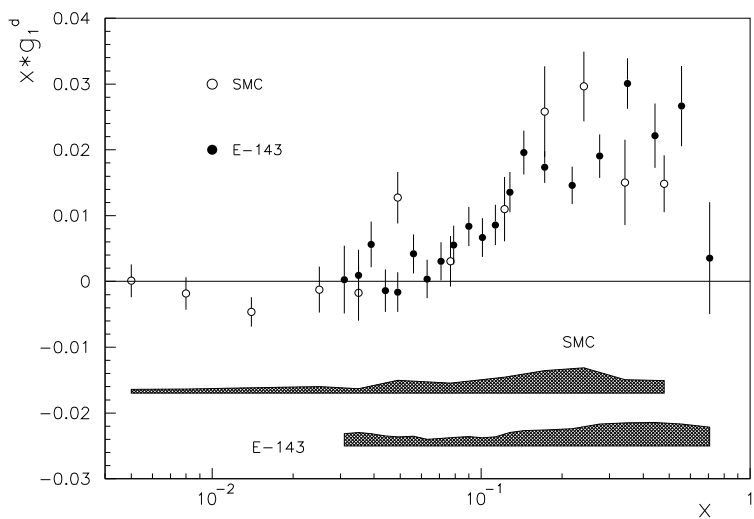,width=10.0cm}
\caption{Data on $x g_1^d(x)$ from SMC~$^{12}$ and E143~$^{15}$.
The E143 data are at $Q^2$= 3~GeV$^2$, while the SMC data are at the
measured $Q^2$ for each $x$ bin.
\label{fig:xg1d}}
\end{figure}

\begin{figure}
\hspace{0.6cm}\psfig{figure=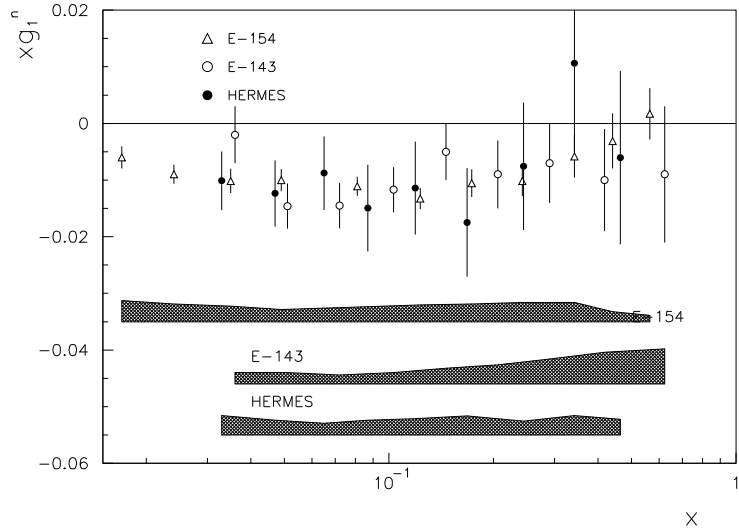,width=9.75cm}
\caption{Data on $x g_1^n(x)$ from E143~$^{15}$, HERMES~$^{17}$, 
and E154~$^{18}$.  The E143 data are at $Q^2$= 3~GeV$^2$, E154 is
at $Q^2$= 5~GeV$^2$, and the HERMES data are at the measured $Q^2$
in each $x$ bin.
\label{fig:xg1n}}
\end{figure}

The EMC results~\cite{as88} published in 1988, which indicated that
 little of the spin of the proton comes from the spins of the quarks,
 generated a series of experiments at CERN (SMC: H \& D targets), SLAC
 (E142 \& 154: $^3$He, and E143 \& 155: H \& D), and DESY (HERMES: H, D, \&
 $^3$He) to study the {\em ``nucleon spin puzzle''}.  As a result, we
 now have precise data on $g_1(x)$ over a range of $Q^2$.  A selection
 of these data, as of the time of this workshop (Feb.'98), is shown in
 Figs.~\ref{fig:xg1p},\ref{fig:xg1d},\ref{fig:xg1n} (see the figure
 captions for references).  There are several choices for the form of
 these plots.  Here, the abscissa is plotted on a logarithmic scale to
 allow the points at low-$x$ to be seen clearly.  $xg_1(x)$ rather
 than $g_1(x)$ is plotted on the ordinate because this is the
 appropriate quantity to plot to illustrate the contribution of each
 data point to the integral of $g_1(x)$, the importance of which in
 terms of the spin content of the nucleon will be discussed in
 Sec.~\ref{sec:sumrules}-\ref{sec:content}.

%
%
%

A few general points can be made about the data.  First, although they
are taken with widely varying experimental techniques (see
Sec.~\ref{sec:exp}), the data are consistent with each other, giving
confidence in the results.  Second, while the $Q^2$ varies substantially
for the different experiments ($\langle Q^2 \rangle$= 2-10~GeV$^2$),
$g_1(x)$ shows relatively little $Q^2$-dependence over most of the
$x$-range, at least within the experimental uncertainties.  Third,
while the SMC data have worse statistical precision than either the
SLAC experiments or HERMES, they extend to lower $x$ which is
important for the extrapolation of $g_1(x)$ to $x=0$.  Finally, the
precision of the data is quite good and the experiments are now close
to being limited by systematic uncertainties.

One should notice that while $g_1^p(x)$ is large and positive for
$0.1<x<0.6$, $g_1^n(x)$ is small and negative for all $x$.  The reader
is referred to the experimental papers for a more complete
presentation of the data.

\section{The Interpretation of $g_1(x)$
in Terms of Sum Rules}\label{sec:sumrules}

The spin structure functions are often interpreted using sum rules
which relate the integrals of $g_1(x)$ over all $x$ to the
polarisation of the quarks in the nucleon.

\subsection{The Bjorken Sum Rule}\label{sec:BJ}

\begin{figure}
\hspace{0.75cm}\psfig{figure=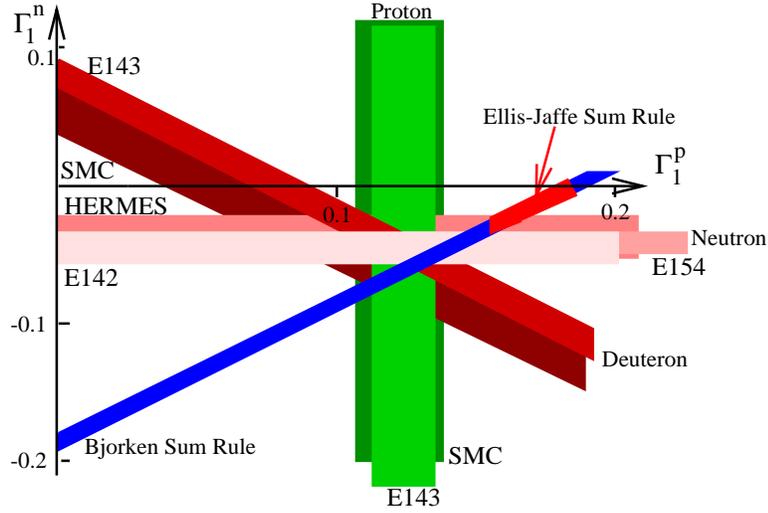,width=4.0in}
\caption{Plot of $\Gamma_1^n$ versus $\Gamma_1^p$.  The predictions of 
the Bjorken and Ellis-Jaffe sum rules are shown on the diagonal band
from the lower left to the upper right of the figure.  While the data
and the Bjorken sum rule overlap within one sigma, the Ellis-Jaffe
prediction is roughly two sigma away from the overlap region in the data.
\label{fig:sumrule}}
\end{figure}

If we use Eqn.~\ref{eqn:g1} for the proton and the neutron, we get:
\begin{eqnarray}
\Gamma_1^p= \int_{0}^{1} g_1^p(x)\,dx= \frac{1}{2} \left( \frac{4}{9} 
\Delta u + \frac{1}{9} \Delta d + \frac{1}{9} \Delta s \right) 
\label{eqn:gamma1} \\
\Gamma_1^n= \int_{0}^{1} g_1^n(x)\,dx= \frac{1}{2} \left( \frac{1}{9} 
\Delta u + \frac{4}{9} \Delta d + \frac{1}{9} \Delta s \right),
\label{eqn:gamma2}
\end{eqnarray}

\noindent
where we have used isospin symmetry to make the simplifications:
\begin{eqnarray}
\Delta u_p = \Delta d_n = \Delta u \\
\Delta d_p = \Delta u_n = \Delta d \\
\Delta s_p = \Delta s_n = \Delta s
\end{eqnarray}

\noindent
with the corresponding equations for the anti-quarks.  $\Delta u$,
$\Delta d$, and $\Delta s$ in Eqns.~\ref{eqn:gamma1},~\ref{eqn:gamma2}
are therefore for the {\em proton}.  The sea cancels in the
difference of the two integrals so that
\begin{equation}
\int_{0}^{1} [g_1^p(x) - g_1^n(x)] \, dx= \frac{1}{6} (\Delta u - \Delta d)=
\frac{1}{6} \cdot \left| \frac{g_A}{g_V} \right|
\label{eqn:bj}
\end{equation}

\noindent
where the final step comes from neutron $\beta$-decay, and $g_A$ and
$g_V$ are the axial vector and vector coupling constants. This sum
rule, first derived by Bjorken~\cite{bj66} using current algebra, has
little model dependence and is fundamental to QCD.  If higher order
QCD effects are taken into account, the full expression for the {\em
Bjorken Sum Rule} is obtained~\cite{la91}:
\begin{eqnarray}
\int_{0}^{1} [g_1^p(x) - g_1^n(x)] \, dx= \frac{1}{6} \cdot 
\left| \frac{g_A}{g_V} \right| \cdot Corr \\ \nonumber \\
Corr= 1 - \left( \frac{\alpha_s(Q^2)}{\pi} \right) 
- 3.5833 \left( \frac{\alpha_s(Q^2)}{\pi} \right)^2 \nonumber \\
- 20.2153 \left( \frac{\alpha_s(Q^2)}{\pi} \right)^3
- {\mathcal{O}}(130) \left( \frac{\alpha_s(Q^2)}{\pi} \right)^4, \nonumber
\end{eqnarray}

\noindent
where $\alpha_s$ is the $Q^2$ dependent strong coupling constant.
The precise data on $g_1^p(x)$, $g_1^n(x)$, and $g_1^d(x)$ can be used
to test the Bjorken sum rule.  The current situation is shown in
Fig.~\ref{fig:sumrule} which is a plot of the neutron integral
($\Gamma_1^n$) versus the proton integral ($\Gamma_1^p$).  The overlap
of the bands for the three targets is consistent with the sum rule to
one standard deviation.  The Bjorken sum rule has been verified to
about 10\%.

\subsection{The Ellis-Jaffe Sum Rules}\label{sec:EJ}

Because of the difficulty of producing a polarised neutron target, it
became clear that data on the proton would be available much before
data on the neutron.  Ellis and Jaffe~\cite{el74} derived sum rules
for $\Gamma_1^p$ and $\Gamma_1^n$ separately using SU(3)$_F$ flavour
symmetry and the assumption that $\Delta$s= 0.  One can write the
following from neutron and hyperon $\beta$-decays:
\begin{eqnarray}
\Delta u - \Delta d = F + D \label{eqn:SU3} \\
\Delta u + \Delta d -2 \Delta s = 3F - D \nonumber
\end{eqnarray}

\noindent
where $F$ and $D$ are SU(3)$_F$ coupling constants.  If $\Delta s= 0$,
we can solve Eqs.~\ref{eqn:SU3} for $\Delta u$ and $\Delta d$ and write
Eqs.~\ref{eqn:gamma1},~\ref{eqn:gamma2} in terms of $F$ and $D$:
\begin{eqnarray}
\int_{}^{} g_1^p(x)~dx= \frac{1}{18} (9F-D) \\
\int_{}^{} g_1^n(x)~dx= \frac{1}{18} (6F-4D)
\end{eqnarray}

\noindent
These sum rules are usually written in a different form:
\begin{eqnarray}
\int_{}^{} g_1^p(x)~dx= \frac{1}{12} \cdot \left| \frac{g_A}{g_V} 
\right| \cdot 
\left(1 + \frac{5}{3} \ \frac{3F-D}{F+D} \right) \times QCD~Corr. \\
\int_{}^{} g_1^n(x)~dx= \frac{1}{12} \cdot \left| \frac{g_A}{g_V} 
\right| \cdot 
\left(-1 + \frac{5}{3} \ \frac{3F-D}{F+D} \right) \times QCD~Corr.
\end{eqnarray}

\begin{figure}
\hspace{0.3cm}\psfig{figure=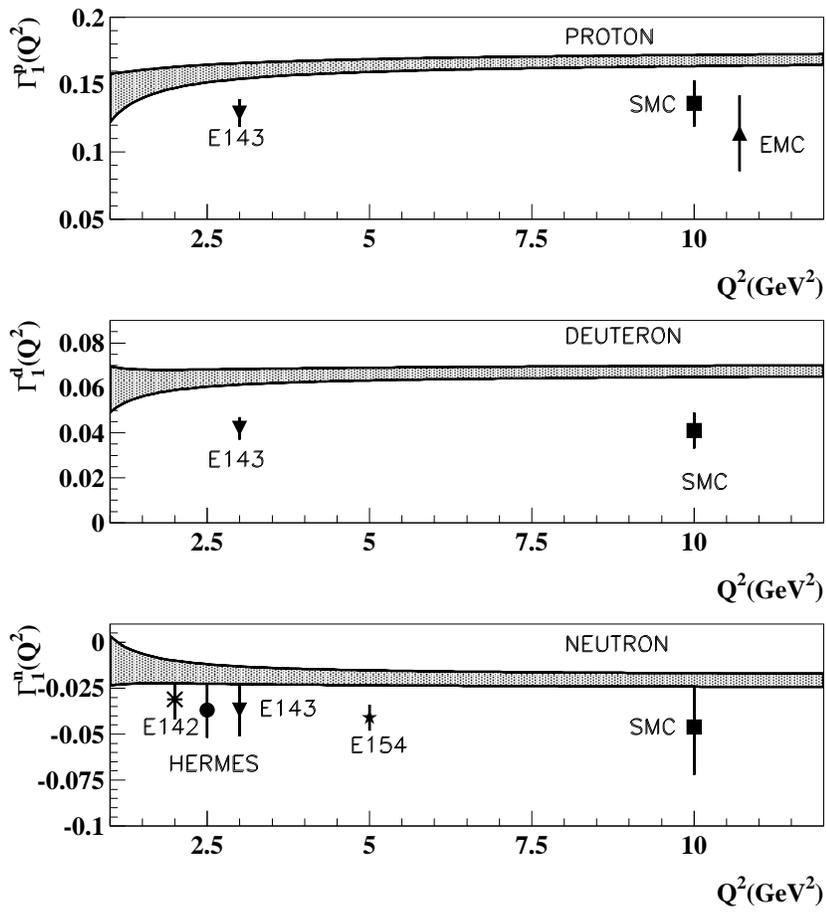,width=11.0cm}
\caption{Results for $\Gamma_1^p$, $\Gamma_1^n$, and $\Gamma_1^d$ compared
to the predictions of Ellis and Jaffe (shaded area).
\label{fig:EJ}}
\end{figure}

\noindent
Using the fact that $F + D= g_A/g_V= 1.257 \pm 0.003$ from neutron
$\beta$-decay and the determination of $F/D$ from a global fit to
hyperon $\beta$-decays ($F/D= 0.575 \pm 0.016$), we can get numerical
predictions for the Ellis-Jaffe integrals.  For example~\cite{cl93},
at $Q^2$= 5~GeV$^2$:
\begin{displaymath}
\Gamma_1^p(EJ)= 0.172 \pm 0.009~;~\Gamma_1^n(EJ)= -0.018 \pm 0.009
\end{displaymath}

\noindent
where higher order QCD corrections are included.  While the Bjorken sum
rule is verified by data, there is significant disagreement 
between the measured values of $\Gamma_1^p$ and $\Gamma_1^n$ and the
Ellis-Jaffe predictions.  The situation is summarized in Fig.~\ref{fig:EJ}
where deviations of several $\sigma$ can be seen between the data and the
predictions.  The values of the Ellis-Jaffe integrals for the different
experiments are given in Table~2.

\begin{table}
\begin{center}
\caption{Values of the Ellis-Jaffe integrals from the various polarised
DIS experiments.  The second number in the fourth column is the
statistical uncertainty on the measurement, while the third number is
the systematic error.}

\ \

\begin{tabular}{|lccrc|}
\hline
Experiment & Integral     & $\langle Q^2 \rangle$ (GeV$^2$) & 
\multicolumn{1}{c}{Measurement} & Ref. \\
\hline
EMC   & $\Gamma_1^p$ & 10.7 &  0.114 $\pm$ 0.012 $\pm$ 0.026 & \cite{as88} \\
\hline
SMC   & $\Gamma_1^p$ &  10  &   0.136 $\pm$ 0.013 $\pm$ 0.011 & \cite{ad97a} \\
      & $\Gamma_1^d$ &  10  &   0.041 $\pm$ 0.006 $\pm$ 0.005 & \cite{ad97b} \\
      & $\Gamma_1^n$ &  10  & --0.046 $\pm$ 0.018 $\pm$ 0.019 & \cite{ad97b} \\
\hline
E142  & $\Gamma_1^n$ &   2  & --0.031 $\pm$ 0.006 $\pm$ 0.009 & \cite{an96}  \\
\hline
E143  & $\Gamma_1^p$ &   3  &   0.127 $\pm$ 0.004 $\pm$ 0.009 & \cite{ab95a} \\
      & $\Gamma_1^d$ &   3  &   0.042 $\pm$ 0.003 $\pm$ 0.004 & \cite{ab95b} \\
      & $\Gamma_1^n$ &   3  & --0.037 $\pm$ 0.008 $\pm$ 0.011 & \cite{ab95b} \\
\hline
E154  & $\Gamma_1^n$ &   5  & --0.041 $\pm$ 0.004 $\pm$ 0.006 & \cite{ab97a} \\
\hline
HERMES & $\Gamma_1^n$ & 2.5  & --0.037 $\pm$ 0.013 $\pm$ 0.008 & \cite{ac97} \\
\hline
\end{tabular}
\end{center}
\label{tab:EJsum}
\end{table}

The disagreement of the data with the predictions of Ellis and Jaffe
is not surprising for several reasons.  From the theory standpoint,
the assumptions of SU(3)$_F$ symmetry and $\Delta s= 0$ are crude
at best. On the experimental side, there is great controversy
over how the extrapolation to the unmeasured region at low
$x$ should be done.  Regge theory has been used in the past but recent
low $x$ data ($F_2$ and $g_1$) have shown this to be inappropriate.
Several other functions have been tried in the low $x$ region but there
is no agreement to date on the proper form to use (see for example 
Ref.~\cite{ab97b} and references therein).

\section{The Spin Content of the Nucleon}\label{sec:content}

The expressions for the integrals $\Gamma_1^p$ and $\Gamma_1^n$, and
for the matrix elements of hyperon $\beta$-decays in terms of $\Delta
u$, $\Delta d$, and $\Delta s$ can be used to deduce the fraction of
the nucleon spin due to the spins of the quarks.  In practice, global
fits are made to all the data.  However as an illustration, a simple
example is shown here which has the added advantage of giving
historical insight into the {\em `Spin Crisis'} since it uses the
original EMC data~\cite{as88}.  Consider the following system of three
equations and three unknowns:
\begin{eqnarray}
\Gamma_1^p= \frac{1}{2} (\frac{4}{9} \Delta u + \frac{1}{9} \Delta d +
\frac{1}{9} \Delta s)= 0.114 \pm 0.029 ~~ (EMC) \nonumber \\
\Delta u - \Delta d= 1.257 \pm 0.011 ~~ ({\rm neutron}~\beta-{\rm decay}) \\
\Delta u + \Delta d - 2 \Delta s= 0.579 \pm 0.016 ~~ ({\rm hyperon} 
~\beta-{\rm decay}). \nonumber
\end{eqnarray}

\noindent
Solving for $\Delta u$, $\Delta d$, and $\Delta s$, we get:
\begin{equation}
\Delta u= 0.705 \pm 0.09 ~~;~~ \Delta d= -0.553 \pm 0.09 ~~;~~
\Delta s= -0.214 \pm 0.09.
\end{equation}

\noindent
The total contribution to the nucleon's spin from the quark spins
is therefore

\begin{equation}
\Delta \Sigma= \Delta u + \Delta d + \Delta s= -0.062 \pm 0.16,
\end{equation}

\noindent
consistent with 0 and several standard deviations from what was
expected.  The EMC did a somewhat different analysis (assuming the
validity of the Bjorken sum rule) but they also obtained $\Delta \Sigma
\approx 0$.  This surprising result lead to the new round of experiments
designed specifically for polarised DIS.  Now that many more data
exist, global fits are done to get the $\Delta q$'s with resulting
values of $\Delta \Sigma$ in the range 0.2-0.3, still significantly
below expectation.

In the naive quark parton model, all the spin of the nucleon comes
from the spin of the quarks so that $\Delta \Sigma$= 1.  If
relativistic wave functions are used

\begin{center}
$\psi= \left[ \begin{array}{c}
\chi \\ \frac{\sigma \cdot p}{E + m} \cdot \chi \end{array} \right]$,
\end{center}

\noindent
the lower components contribute orbital angular momentum and $\Delta
\Sigma$ is reduced to 0.75~\cite{close}.  The Ellis-Jaffe prediction
for $\Delta \Sigma$ can be obtained from Eqn.~\ref{eqn:SU3} with
$\Delta s$=0:
\begin{displaymath}
\Delta u + \Delta d= 3F - D\ \ \ \Rightarrow\ \ \ \Delta \Sigma= 
3 \times 0.459 - 0.791= 0.586.
\end{displaymath}

It is clear that no matter what one does, the measurements of $\Delta
\Sigma$ are all in disagreement with the predictions indicating that
there is still something we do not understand about the spin of the
nucleon; or that there is something wrong with our interpretation of
$\Delta \Sigma$ (see below).

\section{The $Q^2$ Evolution of the Structure Functions}\label{sec:q2}

\subsection{Alternate expression for $g_1(x)$}

Eqns.~\ref{eqn:g1}, \ref{eqn:gamma1}, and \ref{eqn:gamma2} give
simple expressions for $g_1(x)$ in term of the quark distributions.
These can be generalized to include gluons and written in a form which
is convenient for the $Q^2$ evolution of the structure
functions~\cite{al97}:
\begin{equation}
g_1(x,Q^2)= \frac{\langle e^2 \rangle}{2}\ \left[ C_{ns} \otimes 
\Delta q_{ns} + C_s \otimes \Delta \Sigma + 2 n_f\ C_g \otimes 
\Delta g \right],
\label{eqn:g1q2}
\end{equation}
\begin{eqnarray}
{\rm where~} \ \ \langle e^2 \rangle & = & \frac{1}{n_f} \sum_{i=1}^{n_f} 
e_i^2, \\
\Delta q_{ns} & = & \sum_{i=1}^{n_f} \left( \frac{e_i^2}{\langle e^2 \rangle} 
- 1 \right) (\Delta q_i + \Delta \bar{q}_i) \label{eqn:ns} \\
 & = & \frac{3}{5} (\Delta u - \Delta d)\ \ {\rm (for~2~flavours)},
\ \ {\rm ~and} \nonumber \\
\Delta \Sigma & = & \sum_{i=1}^{n_f} (\Delta q_i + \Delta \bar{q}_i).
\end{eqnarray}

\noindent
$\Delta q_{ns}$ is called the {\em non-singlet} quark
distribution, $\Delta \Sigma$ is the {\em singlet} quark
distribution, $\Delta g$ is the gluon polarisation, $n_f$ is the
number of quark flavours, and $C_{ns}$, $C_s$, and $C_g$ are
coefficient functions calculable in QCD.  The $\otimes$ in Eqn.~\ref{eqn:g1q2}
refers to a convolution defined as follows:
\begin{equation}
(C \otimes q)(x,Q^2)=~\int_x^1\,C\left(\frac{x}{z},\alpha_S\right)\,
q(z,Q^2)\ \frac{dz}{z}.
\end{equation}

The advantages of this formulation can be seen if one considers the 
evolution of $g_1$ with $Q^2$.  This evolution is given by the
DGLAP equations~\cite{al77}:
\begin{eqnarray}
\frac{d}{dt}\, \Delta q_{ns}= \frac{\alpha_s(t)}{2 \pi}\ P_{qq}^{ns} \otimes 
\Delta q_{ns}\ \ \ \ ;\ \ \ \ t= log \left( \frac{Q^2}{\Lambda^2}
\right) \label{eqn:q2a} \\
\frac{d}{dt} \left( \begin{array}{c} \Delta \Sigma \\ \Delta g 
\end{array} \right)= 
\frac{\alpha_s(t)}{2 \pi} \left( \begin{array}{lr}
P_{qq}^s & 2 n_f P_{qg}^s \\ P_{gq}^s & P_{gg}^s \end{array} \right) \otimes
\left( \begin{array}{c}  \Delta \Sigma \\ \Delta g \end{array} \right).
\label{eqn:q2b}
\end{eqnarray}

\noindent
$P_{qq}^{ns}$ and the $P_{ij}^s$ are splitting functions and $\Lambda$
is the constant of the strong interaction ($\approx$ 200-300~MeV).
The first thing to notice is that the $Q^2$ evolution of the
non-singlet term does not depend on the gluons.  Since the Bjorken sum
rule is related to the non-singlet term (see Eqns.~\ref{eqn:bj},
\ref{eqn:ns}), it depends only on the quark distributions.  On the
other hand, the evolution of the singlet term depends on both the
quark and gluon distributions. Therefore, we must be very careful in
our interpretation of $\Delta \Sigma$.  In fact, the quark
distributions as measured in DIS ($\Delta \tilde{q}$) are modified by
a gluon term $\Delta g$~\cite{al88}:
\begin{equation}
\Delta \tilde{q}= \Delta q - \frac{\alpha_s n_f}{2 \pi}\ \Delta g.
\end{equation}

\noindent
$n_f$ is the number of active quark flavours (3 if heavy quarks are
ignored).  This is thought to be the best candidate to explain the
relatively small measured value of $\Delta \Sigma$, and the search for
an indication that the gluons are polarised has become a focus of the
field.

\subsection{Gluon Polarisation from $Q^2$ Evolution}

\begin{figure}
\psfig{figure=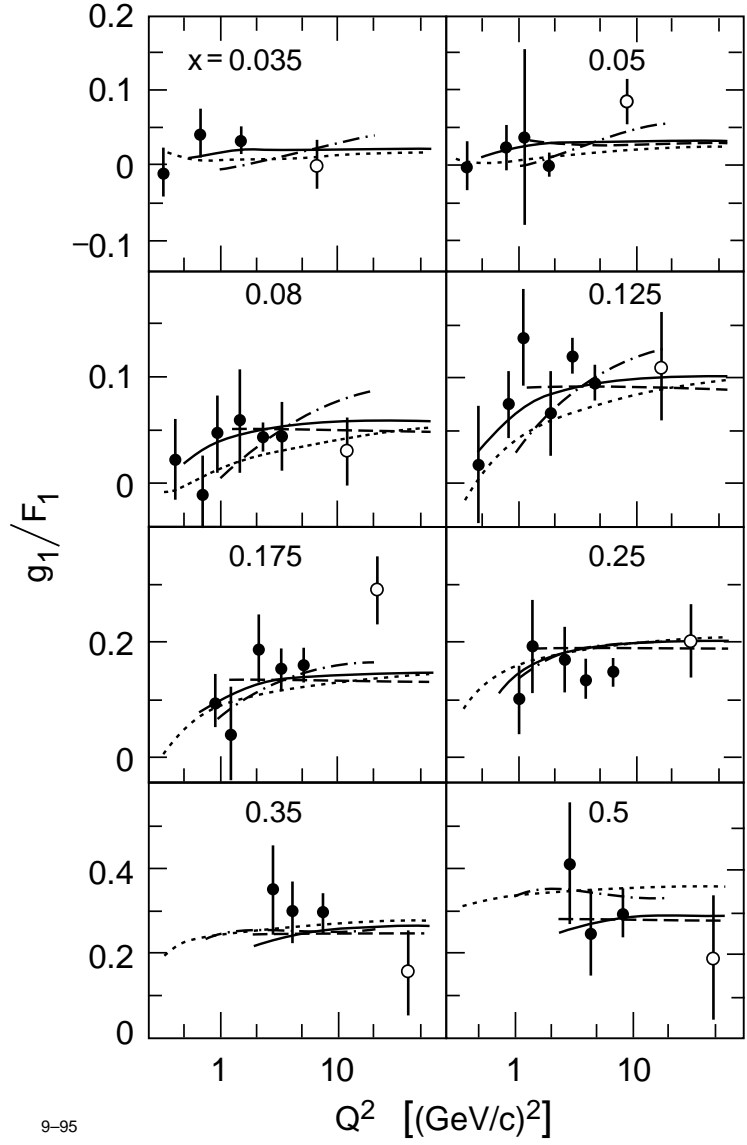,height=6.0in}
\caption{$g_1^d/F_1^d$ as a function of $Q^2$ in separate $x$ bins.
The curves are different global fits to the asymmetries.
This figure is taken from Ref.~$^{27}$ where more 
detail can be found on the fits.
\label{fig:evolv}}
\end{figure}

There have been attempts to extract a value of $\Delta g$ from the
$Q^2$ evolution of $g_1(x)$ by using
Eqns.~\ref{eqn:g1q2},~\ref{eqn:q2a}, and~\ref{eqn:q2b} to perform a
global fit to data at different $Q^2$.  The result of these fits for the
E143 data~\cite{ab95c} is shown in Fig.~\ref{fig:evolv} where some
$Q^2$ dependence is seen at low $Q^2$, albeit small.  Values of
$\Delta g$ in the range 1.5-2.0 have been deduced in several
studies~\cite{al97,gl96,ba95,ge96}.  However, the uncertainties are large
and the data are consistent with $\Delta g$=0.  Much better data will
be required for this technique to produce reliable results.

\section{Other Structure Functions}\label{sec:others}
\subsection{$g_2(x)$}\label{subsec:g2}

Until now the discussion has centered on $g_1$.  However, several
other structure functions can be measured.  $g_2(x)$ has already been
introduced in Eqn.~\ref{eqn:wmunu} and Sec.~\ref{subsec:practice}.
While $g_2$ could be important as a correction to $g_1$, it is also
interesting in its own right because it could offer a way of isolating
the effects of a twist-3 operator~\cite{ja91a}.  Twist is a number
that arises in the operator product expansion and is an indication of
the order of the operator involved~\cite{ma92}.  Twist-2 operators are
leading-order, while twist-3 are next-to-leading order, etc.  The
following expression for $g_2$ shows a term ($\tilde{g}_2$) which is
pure twist-3:
\begin{equation}
g_2(x,Q^2)= g_2^{WW}(x,Q^2) + \tilde{g}_2(x,Q^2)
\end{equation}
\begin{equation}
{\rm with}\ \ \ g_2^{WW}(x,Q^2)= -g_1(x,Q^2) + \int_{x}^{1} 
\frac{g_1(y,Q^2)}{y}\, dy
\end{equation}

\begin{figure}
\hspace{2.0cm}\psfig{figure=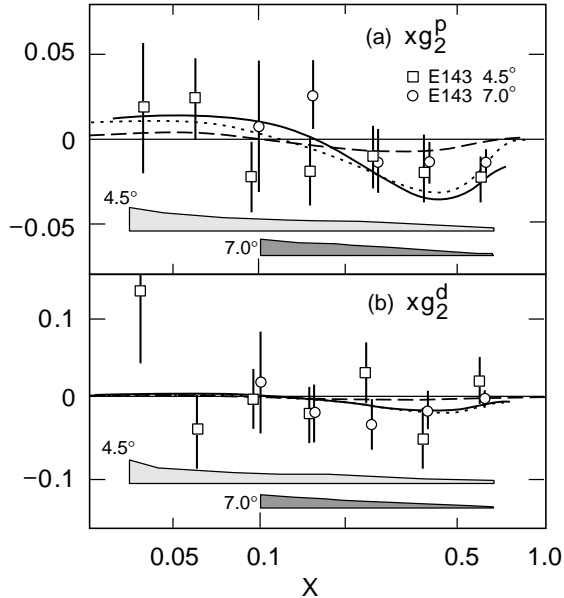,width=3.0in}
\caption{Measurements of (a) $x g_2^p$ and (b) $x g_2^d$ from E143 at 
SLAC~$^{34}$.  Systematic uncertainties are indicated by the bands.
The solid curve shows the $g_2^{WW}$ calculations.  Bag model
calculations at $Q^2$= 5.0~GeV$^2$ by Stratmann~$^{35}$ (dotted), and
Song and McCarthy~$^{36}$ (dashed) are also shown.
\label{fig:g2}}
\end{figure}

\noindent
The term $g_2^{WW}$ was derived by Wandzura and Wilczek~\cite{wa77}.  It
is twist-2 (i.e. leading order) and can be calculated from $g_1$.  A
deviation of $g_2$ from $g_2^{WW}$ would signal the effect of a
twist-3 operator.  Data on $g_2$ from SLAC~\cite{ab96} are shown in
Fig.~\ref{fig:g2}.  The data are consistent with either 0 or
$g_2^{WW}$ which is shown as the solid line.  They are not precise
enough to distinguish between these two possibilities or to isolate
any possible twist-3 contribution.  However, these data confirm that
corrections to $g_1$ due to $g_2$ are small.  More precise data on
$g_2$ are expected from SLAC in 1999.

\noindent
Burkhart and Cottingham~\cite{bu70} have derived a sum rule for $g_2$:
\begin{equation}
\int_{0}^{1} g_2(x)\, dx= 0\ \ \ \ \ (Q^2 \rightarrow \infty).
\end{equation}

\noindent
The data in Fig.~\ref{fig:g2} have been integrated with the following
results~\cite{ab96}:
\begin{equation}
\int_{0.03}^{1} g_2^p(x)\, dx= -0.013 \pm 0.028
\end{equation}
\begin{equation}
\int_{0.03}^{1} g_2^d(x)\, dx= -0.033 \pm 0.082
\end{equation}

\noindent
both consistent with 0.

\subsection{Still more Structure Functions}

\subsubsection{$h_1(x)$}

$h_1(x)$ is a twist-2 structure function like $F_1(x)$ and
$g_1(x)$~\cite{ja91b}.  It is related to tensor currents in the
nucleon.  However, it is more difficult to measure because it is
chiral-odd.  A longitudinally polarised beam and a transversely
polarised target are needed to measure $h_1(x)$.  Furthermore, because
this structure function is chiral-odd, pions must be detected in
coincidence with the lepton to conserve chirality.

\subsubsection{$b_1(x)$}

$b_1(x)$ is measured with an unpolarised beam and a longitudinaly polarised
D target in each of three substates ($m_d$= 1, 0, --1).  It is given by the
expression:
\begin{equation}
b_1(x)= \frac{1}{2} \sum_{f}e^2_f~[q^+_{f}(x)+q^-_{f}
(x)-2q^0_f(x)]
\end{equation}

\noindent
and measures quarks from nuclear binding~\cite{ho89}.

\subsubsection{$\Delta(x)$}

$\Delta(x)$ is also measured with an unpolarised beam and a D target, but
in this case the target is transversely polarised in the $m_d$=0 state.  The
$\phi$ distribution of the scattered leptons determines $\Delta(x)$
which is sensitive to gluons not identified with individual 
nucleons~\cite{ja89}.

\section{Experiments to Study the Spin Structure of the Nucleon}
\label{sec:exp}

We will now spend some time describing the recent experiments which
have studied polarised deep inelastic scattering.  The emphasis will
be on the unique features of each one.

\subsection{CERN: The Spin Muon Collaboration (SMC)}

\begin{figure}
\psfig{figure=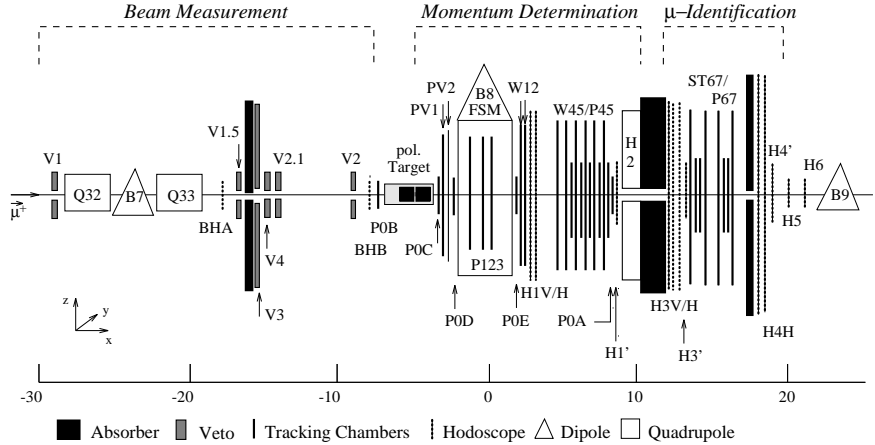,width=12.0cm}
\caption{SMC Spectrometer.
\label{fig:SMCspect}}
\end{figure}

In the SMC experiment, 100-190~GeV muons scatter from butanol
($\rm C_4\vec{H}_9O\vec{H}$), deuterated butanol ($\rm C_4\vec{D}_9O\vec{D}$),
and ammonia ($\rm N\vec{H}_3$) targets (see Ref.~\cite{ad97a} and references
therein).  The experiment is an extension of the pioneering work done
at CERN in the eighties and uses essentially the same spectrometer as
the EMC, shown in Fig.~\ref{fig:SMCspect}.  This device is centered
around a large aperture dipole magnet which provides a 2.3~T.m
magnetic field.  Particle tracks are determined by a series of
multi-wire proportional chambers and drift chambers.  Muon
identification is done using a thick hadron absorber behind the
tracking chambers, followed by streamer tubes.  The trigger is
generated by three hodoscopes behind the absorber.

The muon beam is produced by the decay in flight of pions and kaons.
The current is equivalent to about 0.5~pA ($4.5 \times 10^7$ muons
per spill in 2.4~s every 14.4~s).  The muons are polarised because of
parity violation in the pion and kaon decays.  The typical beam
polarisation was 80\%, determined by a calculation and confirmed by a
measurement of the Michel spectrum of $\mu$ decays downstream of the
experiment.

The target is polarised using dynamic nuclear polarisation.  In this
technique, the high polarisation of electrons in paramagnetic
impurities is transferred to protons using simultaneous flips of these
spins induced by microwave irradiation.  The polarisation direction is
chosen by adjusting the frequency of the microwaves.  The target
operates in a high magnetic field (2.5~T) at a temperature of 50~mK.
The target orientation can be chosen longitudinal or transverse
to the incoming muon.  One of the remarkable features of the
experiment is that two targets with opposite spin direction are used
simultaneously.  Each target cell is 60~cm long and 5~cm in diameter.
The use of two targets is ideal for measuring yield ratios and
asymmetries, greatly reducing the systematic uncertainties related to
luminosity and detector performance.  On the other hand, the SMC
targets have the disadvantage of containing a large fraction of
unpolarised nucleons which greatly dilutes the experimentally observed
asymmetry.  A measure of this effect is given by the {\em dilution
factor} $f_D$ which is the ratio of polarisable nucleons divided by the
total number of nucleons in the target.  The factor $f_D$ varies from
0.13 to 0.23 in the SMC experiment.  The polarisation of each half of
the target is flipped from time to time to take into account the
difference in the acceptance of the spectrometer for each target
section.  The polarisation is measured using NMR and was found to be
$\approx$~86\% for protons and 25-50\% for deuterons.  Target
thicknesses of about 5~.~10$^{24}$~cm$^{-2}$ were obtained leading to
luminosities on the order of 1-2~.~10$^{31}$~cm$^{-2}$~s$^{-1}$.  The
running conditions for SMC from 1992 to 1996 are summarised in
Table~\ref{tab:smc}.

\begin{table}
\begin{center}
\caption{Running conditions for SMC.}

\ \

\begin{tabular}{cccccl}
Year & E$_B$ (GeV) & p$_B$ & p$_T$ &  $f_D$ & Target \\
\hline
 '92 &    100      &  0.84 &  0.35 & 0.23 & d-butanol \\
 '93 &    190      &  0.79 &  0.86 & 0.13 & p-butanol \\
 '94 &    190      &  0.79 &  0.50 & 0.23 & d-butanol \\
 '95 &    190      &  0.79 &  0.50 & 0.23 & d-butanol \\
 '96 &    190      &  0.79 &  0.89 & 0.17 & p-NH$_3$
\end{tabular}
\label{tab:smc}
\end{center}
\end{table}

Another unique feature of SMC is that it has the highest beam energy
of the polarised DIS experiments and therefore the highest $Q^2$
($\langle Q^2 \rangle \approx$ 10~GeV$^2$) and lowest $x$ ($x_{min}
\approx~0.003$).  The low $x$ points are especially valuable in
determining the best functional form to use for the extrapolation of
$g_1(x)$ to $x=0$.

\subsection{SLAC: E142 - E143 - E154 - E155}

The SLAC experiments use spectrometers in end-station A.  While the
technology used for the polarised electron beam is the same for all
these experiments, the targets are very different (see references
in~\cite{an96} for more details).

The beam is produced by photoemission from an AlGaAs crystal induced by
circularly polarised laser light.  The direction of the electron spin
is chosen at random on a pulse by pulse basis.  The beam polarisation
is measured by M$\o$ller scattering from thin ferromagnetic foils.  It
was about 36\% for E142 and improved dramatically for the subsequent
experiments to 82-86\% by using a strained AlGaAs crystal.  The beam
energy varied from 9.7-29.1~GeV for E142-143 to 48.3~GeV for E154.
Beam currents also varied from 0.5-3.5~$\mu$A (e.g. 2.10$^{11}$ e$^-$
per pulse with a 120~Hz repetition rate for E142, to 3-9.10$^{10}$
e$^-$ per pulse for E154).

E142 and E154 used a $\rm ^3\vec{He}$ target~\cite{an96}, while E143
and E155 used $\rm N\vec{H}_3$, $\rm N\vec{D}_3$ and $\rm \vec{LiD}$
targets.  As discussed in Sec.~\ref{subsec:3he}, any spin effects
observed in $\rm ^3\vec{He}$ scattering can be ascribed to the
neutron.  The SLAC target uses optical pumping by infrared lasers of
Rb atoms, followed by spin-exchange with the $^3$He nuclei.  Two cells
are used, one for optical pumping and one for the scattering,
connected by a thin tube.  This helps reduce the amount of Rb in the
beam which in turn keeps the dilution factor $f_D$ relatively high
($f_D \approx$ 0.33 for E142).  The target cell was 30~cm long by 2~cm
in diameter and contained between 8.4~atm (E142) and 9.5~atm (E154) of
gas at 20$^{\circ}$C.  The polarisation, measured using NMR, was
33\% for E142 and 38\% for E154.  The target polarisation was reversed
from time to time to reduce the systematic uncertainty in the
asymmetry.  The $\rm N\vec{H}_3$ and $\rm N\vec{D}_3$ targets used by
E143 and E155 ($\rm \vec{LiD}$ instead of $\rm N\vec{D}_3$) are
similar to the SMC target in that they also use dynamic nuclear
polarisation.  However, in this case only one target cell is used and
it is much smaller than the SMC target (3~cm long and 2.5~cm in
diameter).  The target field is 4.8~T and the temperature is 1.6~K.
The polarisations were in the range 65-80\% for the proton and
$\approx$25\% for the deuteron.  The dilution factors range from 0.17
(H) to 0.25 (D).  The advantage for E155 of $\rm \vec{LiD}$ over $\rm
N\vec{D}_3$ comes from the fact that the Li, which can be described
approximately as an $\alpha$-particle plus a deuteron, is also
polarised.  Hence the dilution factor is better for $\rm \vec{LiD}$.

\begin{figure}
\begin{turn}{-90}
\psfig{figure=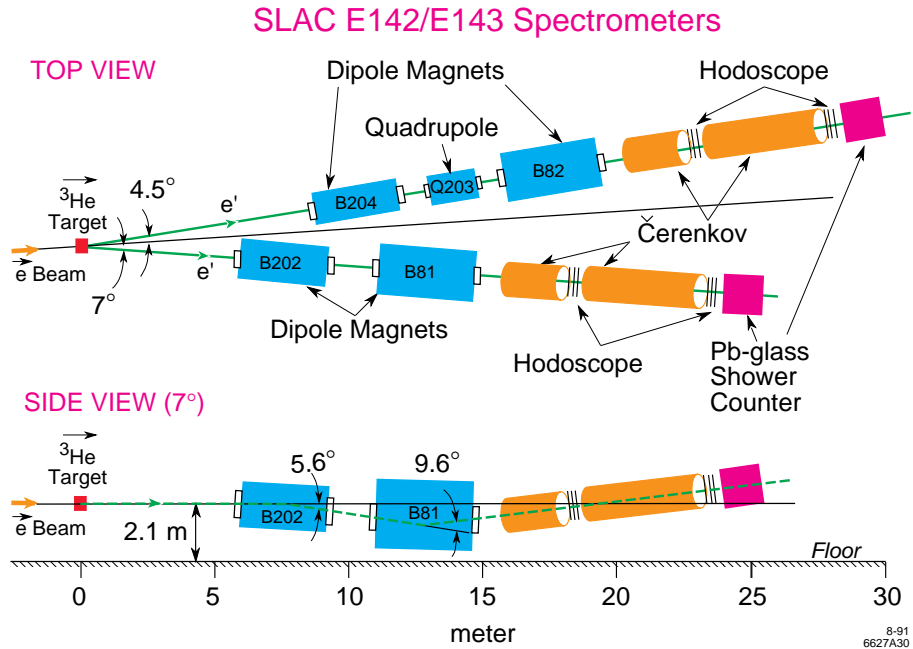,height=12.0cm}
\end{turn}
\caption{Spectrometer for SLAC Experiments E142 and E143.
\label{fig:e143-spect}}
\end{figure}

The spectrometers used for E142-143 are shown schematically in
Fig.\ref{fig:e143-spect}.  The electrons are detected at scattering
angles of 4.5$^{\circ}$ and 7.0$^{\circ}$.  The spectrometers were
modified for operation at higher energy for E154-155.  In particular,
the scattering angles were changed to 2.75$^{\circ}$ and 5.5$^{\circ}$
and a new 10$^{\circ}$ spectrometer was built.  Electrons are
identified by \v{C}erenkov detectors and lead-glass calorimeters while
tracking is done with scintillator hodoscopes.  The geometrical
acceptance of these spectrometers is small compared to the other
experiments.

The running conditions for the SLAC experiments are summarized in
Table~\ref{tab:slac}.

\begin{table}
\begin{center}
\caption{Running conditions for the SLAC experiments.}

\ \

\begin{tabular}{cccccc}
Experiment & E$_B$ (GeV) & p$_B$ & p$_T$   &  $f_D$     & Target \\
\hline
   E142    &    19-26    &  0.36 & 0.33    & 0.33     & $^3$He \\
   E143    &   9.7-29.1  &  0.85 & 0.7 (H) & 0.17 (H) & NH$_3$ \\
           &             &       & 0.25(D) & 0.25 (D) & ND$_3$ \\
   E154    &    48.3     &  0.82 & 0.38    & 0.55     & $^3$He \\
   E155    &    48.3     &  0.8  & 0.83(H) & 0.23     & NH$_3$ \\
           &             &       & 0.23(D) & 0.23     & LiD
\end{tabular}
\label{tab:slac}
\end{center}
\end{table}

\subsection{DESY: HERMES}

The HERMES experiment uses a polarised positron (electron) beam and
pure gas targets internal to the HERA storage ring.  The beam energy
is 27.5~GeV and typical currents range from 40~mA at the start of a
machine fill to 10-12~mA when the beam is dumped.  The positrons are
polarised after acceleration using the Sokolov-Ternov
effect~\cite{so64}.  This takes advantage of a small asymmetry in the
spin-flip amplitude in the emission of synchrotron radiation.
Polarisation builds up over a period of roughly half an hour to an
equilibrium level between 55\% and 60\%.  The polarisation which is
transverse to the beam momentum is rotated into the longitudinal
direction in the HERMES straight section by a series of magnets.  It
is rotated back to transverse downstream of the experiment.  The beam
polarisation is measured using Compton backscattering of circularly
polarised laser light~\cite{ba93}.  HERA is an e-p collider and while
the proton beam is not used by HERMES, it nonetheless goes through the
experiment, displaced horizontally by 72~cm from the positron beam.

One of the strengths of HERMES is the purity of the targets.  The
dilution factor is 1 (the maximum of 1/3 for $^3$He) so that the
experimentally observed asymmetry is not quenched.  Polarised gases
(H, D, $^3$He) flow into a windowless storage cell in the beam pipe.
The cell has an elliptical cross-section (9.8~mm high $\times$ 29~mm
wide) and is 400~mm long.  It increases the areal density of the
target by two orders of magnitude compared to a free atomic beam.  Gas
escapes out of the ends of the cell and is pumped away by a high speed
differential pumping system.  The $\rm \vec{^3He}$ target is polarised
by optical pumping~\cite{ds98}.  However, in contrast to the SLAC
target, the $^3$He is pumped directly so there are no extraneous
atomic species to dilute the target.  The $^3$He nuclei are polarised
by spin exchange collisions with $^3S$ metastable $^3$He atoms in a
glass cell.  The $^3S$ atoms are produced by a weak RF discharge and
polarised by optical pumping with 1083~nm laser light.  The target
thickness is limited to 10$^{15}$ nucleons/cm$^2$ by the requirement
that the lifetime of the positron beam not be affected significantly
by the target.  The polarisation is measured by optical means and
averaged 50\% in 1995.  The $\rm \vec{H}$ and $\rm \vec{D}$ targets
are produced by an atomic beam source (ABS) based on the Stern-Gerlach
effect to select specific spin states.  The thickness of the H target
was $\approx$7.10$^{13}$ atoms/cm$^2$.  The polarisation is measured
by sampling the gas in the cell with a Breit-Rabi polarimeter which
determines the fraction of atoms in each substate.  Polarisations as
high as 92\% were measured.  However, recombination of atoms into
molecules and depolarisation by collisions with the cell walls reduces
this to $\approx$88\% when averaged over all atoms in the cell.

\begin{figure}
\psfig{figure=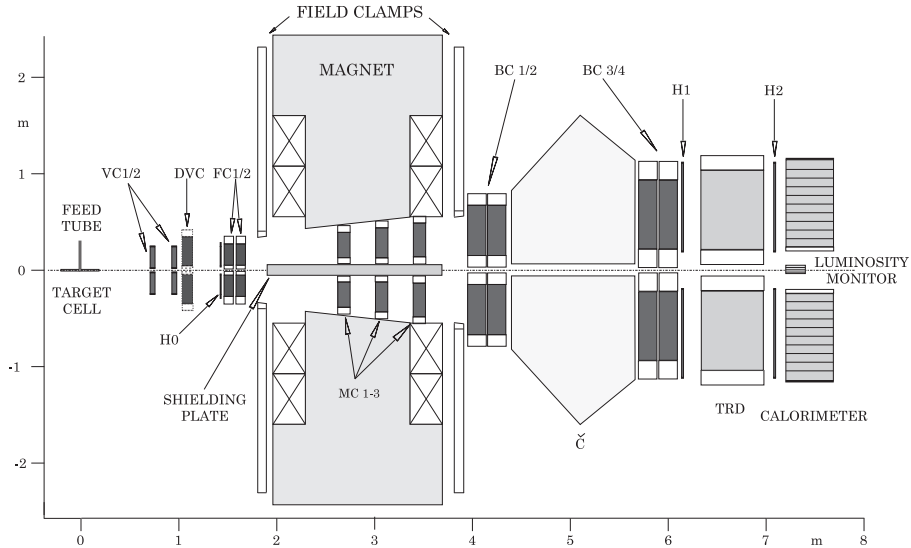,width=12.0cm}
\caption{HERMES Spectrometer.
\label{fig:hermes-spect}}
\end{figure}

The HERMES detector~\cite{ac98} is a forward spectrometer with a large
aperture dipole magnet which provides a field of 1.3~T.m (see
Fig.~\ref{fig:hermes-spect}).  A horizontal iron plate shields the
beams from the magnetic field.  The spectrometer is therefore divided
into two identical halves.  Micro-strip gas chambers just downstream
of the target exit window and drift chambers before and after the
magnet measure the scattering angle and momentum of charged particles.
Proportional chambers inside the magnet track low momentum particles
that do not reach the rear of the spectrometer.  This is particularly
valuable for pions from
$\Lambda$ decays.  Particle identification is provided by
a lead-glass calorimeter preceded by a pre-shower detector, a
six-module transition radiation detector, and a threshold \v{C}erenkov
detector which is used primarily for pion identification.  The
\v{C}erenkov detector has been upgraded to a dual radiator ($\rm
C_4F_{10}$/aerogel) RICH during the 1997-98 shutdown.  This will allow
$\pi$-K-p separation over almost all of the kinematic range of the
experiment (2-20~GeV).  The large acceptance of the spectrometer makes
it possible to detect hadrons in coincidence with the scattered
positron allowing HERMES to make the semi-inclusive measurements
described in section \ref{sec:semiinc} which are the cornerstone
of the HERMES physics program.

\begin{table}
\begin{center}
\caption{Running conditions for HERMES.}

\ \

\begin{tabular}{cccccl}
Year     & E$_B$ (GeV) & p$_B$ & p$_T$ &  $f_D$ & Target \\
\hline
 '95     &    27.5     &  0.5      &  0.50      &  1/3 & $^3$He \\
 '96     &    27.5     &  0.55     &  0.80      &   1  &   H    \\
 '97     &    27.5     &  0.55     &  0.88      &   1  &   H    \\
 '98$^*$ &    27.5     &  0.55$^*$ &  0.85$^*$  &   1  &   D    \\
 '99$^*$ &    27.5     &  0.55$^*$ &  0.85$^*$  &   1  &   D    \\
$^*$ Planned &         &       &       &      &
\end{tabular}
\end{center}
\label{tab:herm}
\end{table}

The running conditions at HERMES are summarized in Table~5.

\section{Semi-inclusive Deep Inelastic Scattering}\label{sec:semiinc}
\subsection{The Next Step towards Understanding the Spin of the Nucleon}

The experimental activity of the last decade has produced precise data
for {\em inclusive} polarised DIS and can be summarized by the following
statements:

\begin{itemize}

\item The EMC result for the integral of $g_1^p(x)$ has been confirmed.

\item The Bjorken sum rule has been verified to about 10\%.

\item The integrals of $g_1^p(x)$, $g_1^n(x)$, and $g_1^d(x)$ disagree
with the Ellis-Jaffe predictions by up to several standard deviations.

\item The quark spins could contribute as little as 20\% of the total
nucleon spin.

\item The nucleon spin puzzle is not yet solved.

\end{itemize}

\noindent
The spin of the nucleon can be broken down into four components:
\begin{equation}
S_z= \frac{1}{2} \left( \Delta V + \Delta S \right) + \Delta g + L_z
\label{eqn:decomp}
\end{equation}

\noindent
where $\Delta V$ is the contribution from the valence quarks, $\Delta
S$ comes from the sea quarks, $\Delta g$ is the gluon polarisation,
and $L_z$ is a possible contribution from the orbital angular momentum
of the partons.  Inclusive polarised DIS is sensitive mainly to $(\Delta
V + \Delta S)$.  Since the quality of the inclusive data sets is now
very good, further progress towards a solution to the spin problem
must come from qualitatively different data, in particular hadrons
produced in coincidence with the scattered lepton.  Some information
on $\Delta g$ can be obtained from the $Q^2$ evolution of $g_1(x,Q^2)$
but we have seen that these effects are small and that the current
data are not precise enough to lead to a reliable result.

The strength of the semi-inclusive measurements is that the type of
hadron detected gives information on the flavour of the struck
quark~\cite{fr89,cl91}.  This is called {\em flavour tagging}.  For
example, a high energy $\pi^+$ coming from a $e^+-p$ collision was
likely produced by the fragmentation of a $u$ quark.  This allows the
spin distributions of particular quark flavours to be determined and
terms in Eqn.~\ref{eqn:decomp} to be isolated.  Some examples of this
are discussed in later sections.  

\noindent
Briefly:

\begin{description}
\item[$\Delta V$:] The valence quark distributions can be derived from
data on the charge dependent polarisation asymmetry in pion 
production (Sec.~\ref{subsec:DV}).
\item[$\Delta S$:]  The light sea quark distributions can be determined 
from the polarisation asymmetry of $\pi^-$ production.  $\pi^-$ is a
$d \bar{u}$ bound state.  Since the charge of the $\bar{u}$ is twice
that of the $d$, it dominates the cross-section (remember the $e_q^2$
term) and this process is sensitive to the light sea.  Information on
the strange sea can be obtained from $K^-$ production since it is a
$\bar{u} s$ state (Sec.~\ref{subsec:DV}).
\item[$\Delta g$:]  The photon-gluon fusion process dominates the 
production of charm particles in DIS.  Information on $\Delta g$ can
therefore be obtained from the polarisation asymmetry in $J/\psi$ and
$D^0$ production (Sec.~\ref{subsec:DG}).
\item[$L_z$:]  It has been proposed that the orbital angular momentum
contribution to the nucleon's spin might be deduced from an azimuthal
asymmetry in hadron production with a transversely polarised target.
However, this is still controversial.
\end{description}

\subsection{Kinematics; the Current and Target Regions}

Two new variables are used to characterize hadrons as coming either
from the struck quark ({\em current region}) or from the target
remnant ({\em target region}):
\begin{equation}
z= \frac{E_h}{\nu}\ \ \ \ ;\ \ \ \ x_f= \frac{2 p_\parallel}{W}.
\end{equation}

\noindent
$z$ is the fraction of the energy transferred in the DIS process
($\nu$) carried by the hadron, and $x_F$ ($x$-Feynman) is a measure of
the longitudinal momentum of the hadron in the $\gamma^*-p$ reference
frame.  Current quarks tend to produce hadrons at high~$z$ and
positive $x_F$, while the target region is characterized by low~$z$,
negative-$x_F$ hadrons.  We will see later that we are interested 
primarily in hadrons produced by current quarks.

The question of how high in $z$ is enough to ensure that the hadron
comes from the current quark is not trivial.  Surely $z>0.5$ is safe
but most hadrons are produced below $z=0.5$ (see Fig.~\ref{fig:had-z})
and lowering the cut on $z$ is necessary to obtain a sufficient number
of events.  This problem was studied by Berger who defined criteria to
ensure that a given hadron came from the current quark~\cite{be87}.
He considered the rapidity:

\begin{figure}
\hspace{1.75cm}\psfig{figure=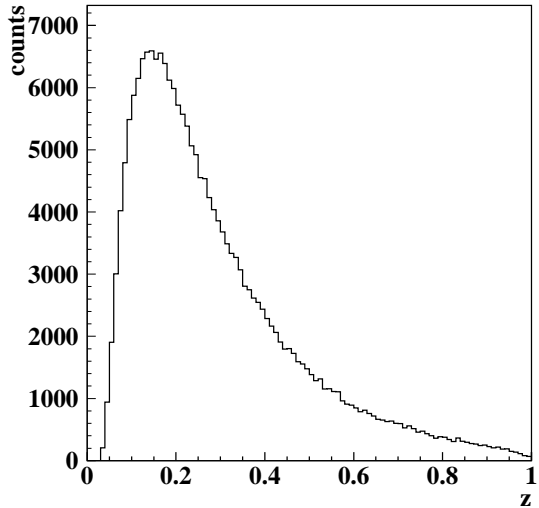,width=3.0in}
\caption{$z$-distribution of hadrons at HERMES in 1996.
\label{fig:had-z}}
\end{figure}

\begin{equation}
y_h= \frac{1}{2} \log \left( \frac{E_h + p_{h,\parallel}}
{E_h - p_{h,\parallel}} \right)
\end{equation}

\noindent
and used the experimentally observed fact that the typical hadronic
correlation length in a jet is 2 units of rapidity.
In other words, it is highly unlikely that two jets separated by
$\Delta y_h~\ge4$ will contain hadrons coming from each other.  Since
the maximum range in $y_h$ which is kinematically allowed for a jet is
\begin{equation}
Y= \log\, W^2= \log \left( Q^2\: \frac{1-x}{x} \right),
\end{equation}

\noindent
the condition $Y \ge 4$ translates to $W \ge 7.4~{\rm GeV}$.
Therefore, if we look along the direction of the virtual photon and
apply Berger's criterion, we are reasonably assured that we are looking
only at hadrons from current quarks.  However, it is important to
realise that this criterion need only be satisfied if we want to
consider {\em all values} of $z$.  The condition on $W$ can be relaxed
if a cut on $z$ is used.  For example, for $Y \ge 2$ (i.e. $W \ge
3~{\rm GeV}$) it is likely that the hadron which is carrying the most
energy (called the {\em leading hadron}) comes from the current
quark.  Another useful guideline is that for $z \ge 0.2$, we need
only require $W \ge 4.8~{\rm GeV}$.

In practice, the best way to be sure that contamination from the target
region is not significant is to determine the quantity of interest
with varying cuts on $z$ and show that the result is stable.

\subsection{Factorisation and the Semi-inclusive Cross-Section}

The multiplicity of hadron $h$ in the final state is given by:
\begin{equation}
\frac{1}{\sigma_{tot}}\ \frac{d^3 \sigma_h}{dx\, dQ^2 dz}= 
\frac{\sum_{i} e_i^2\,\ q_i(x,Q^2)\,\ D_i^h(z,Q^2)}{\sum_{i} e_i^2\,\ 
q_i(x,Q^2)}
\label{eqn:semi}
\end{equation}

\noindent
where the sum is over quarks and anti-quarks separately.
$\sigma_{tot}$ is the inclusive cross-section, $e_i$ is the
charge of the struck quark, $q_i(x,Q^2$) are the quark distribution
functions, and $D_i^h(z,Q^2)$ are the fragmentation functions.
$D_i^h(z,Q^2)$ is the probability density that a quark $i$ will
produce hadron $h$ with energy $z \nu$ when it undergoes the process
of fragmentation as it leaves the nucleon~\cite{fi77}.  In other
words, Eqn.~\ref{eqn:semi} tells us that the mean number of hadrons $h$
produced is a convolution of the probability to strike a quark $q_i$
and the likelihood that this quark will fragment into hadron $h$.  In
principle, the fragmentation functions depend on $Q^2$ as well as $z$
but this is a relatively small effect and is ignored in most of what
follows.  $Q^2$ effects in fragmentation are studied as corrections to
the final results.

Implicit in Eqn.~\ref{eqn:semi} is the concept of {\em factorisation}.
This means that the fragmentation process is independent of the hard
process that produced the outgoing quark, which in DIS is the
absorption of the virtual photon by the struck quark.  In practice,
this means that in Eqn.~\ref{eqn:semi} the fragmentation functions depend
only on $z$.  This leads to the remarkable fact that the fragmentation
functions measured in $e^+-e^-$ collisions should be the same as those
measured in DIS.  This has been verified experimentally~\cite{bi95}.
Also assumed in Eqn.~\ref{eqn:semi} is that the hadrons observed are
from current quarks.  If the latter condition is not met then
factorisation might appear to be broken.

Ultimately, the best way to be sure that factorisation holds is to
plot an $x$-dependent quantity as a function of $z$ and show that it
is independent of this variable.  The same can be done for $z$-dependent
functions in terms of their $x$-dependence.

Some simplifying assumptions can be used to reduce the number of terms in
Eqn.~\ref{eqn:semi}.  For example, consider $\pi^\pm$ production on the
proton.  Even if only three quark flavours ($u$, $d$, and $s$) are included,
we have 12 fragmentation functions.  Using isospin symmetry and charge
conjugation invariance, we can make the following simplifications:
\begin{equation}
D_1= D_u^{\pi^+}= D_{\bar{d}}^{\pi^+}= D_d^{\pi^-}= D_{\bar{u}}^{\pi^-}
\ \ \ \ \ (\rm Favoured~FF)
\label{eqn:favFF}
\end{equation}
\begin{equation}
D_2= D_d^{\pi^+}= D_{\bar{u}}^{\pi^+}= D_u^{\pi^-}= D_{\bar{d}}^{\pi^-}
\ \ \ \ \ (\rm Unfavoured~FF)
\label{eqn:unfavFF}
\end{equation}
\begin{equation}
D_3= D_s^{\pi^+}= D_{\bar{s}}^{\pi^+}= D_s^{\pi^-}= D_{\bar{s}}^{\pi^-}
\ \ \ \ \ (\rm Strange~FF),
\label{eqn:strangFF}
\end{equation}

\begin{figure}
\hspace{1.0cm}\psfig{figure=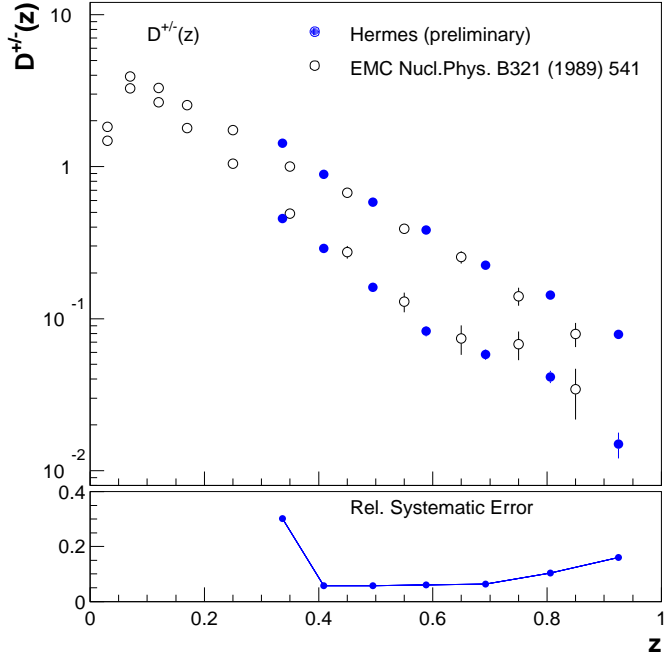,width=3.5in}
\caption{Favoured ($D_1$) and unfavoured ($D_2$) fragmentation functions 
from HERMES$^{50}$ (filled circles) and EMC$^{51}$ (open circles).
The EMC data have been evolved to the same $Q^2$ as HERMES (2.35~GeV$^2$).
\label{fig:FF}}
\end{figure}
\begin{figure}
\hspace{0.5cm}\psfig{figure=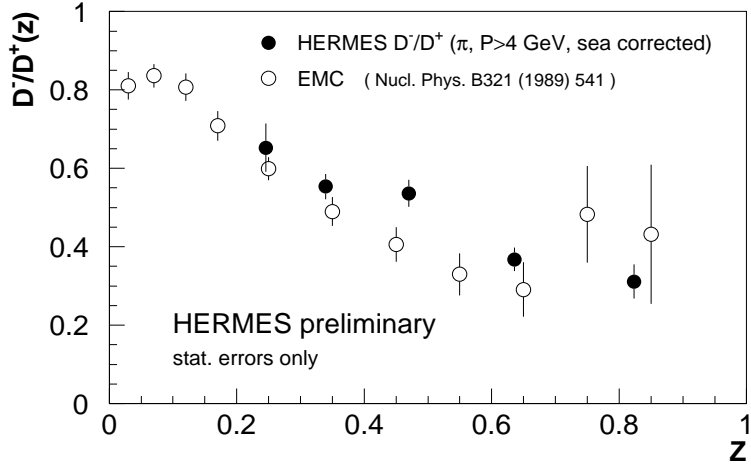,width=4.0in}
\caption{Plot of the ratio of the favoured and unfavoured fragmentation
functions which qualitatively shows the behaviour predicted by 
Eqn.~\ref{eqn:indFF}.
\label{fig:d2-d1}}
\end{figure}


\noindent
reducing the number of fragmentation functions to 3.
Eqn.~\ref{eqn:favFF} states that the probability of a $u$ or a
$\bar{d}$~quark to produce a $\pi^+$ is the same as for a $d$ or a
$\bar{u}$~quark to produce a $\pi^-$.  Since these are valence quarks
in the final state hadron, this fragmentation function ($D_1$) is
called {\em favoured}.  The processes referred to in
Eqn.~\ref{eqn:unfavFF} are more complicated and therefore $D_2$ should
be smaller than $D_1$.  $D_2$ is called the {\em unfavoured}
fragmentation function.  Plots of $D_1(z)$ and $D_2(z)$ from the
HERMES~\cite{ge98} and EMC~\cite{ar89} experiments are shown in
Fig.~\ref{fig:FF} where it can be seen that $D_1$ is several times as
large as $D_2$ in the mid-$z$ range.
Finally $D_3$ is the strange fragmentation function and should be
similar in size to $D_2$.  Fragmentation (or hadronisation) is
governed by long-range dynamics.  It is therefore non-perturbative and
we must use phenomenological models to describe it.  One such simple
model is the independent fragmentation model~\cite{fi77} which leads to:
\begin{equation}
\frac{D_2(z)}{D_1(z)}= \frac{1-z}{1+z}.
\label{eqn:indFF}
\end{equation}
Eqn.~\ref{eqn:indFF} predicts $D_1$=$D_2$ at $z=0$ where the target
region dominates and the hadronisation process is complicated enough
that $\pi^-$'s should be produced in the same amount as $\pi^+$'s.
Eqn.~\ref{eqn:indFF} also predicts $D_2$=0 at $z=1$.  This is
intuitively correct since a hadron carrying all of the energy
transferred in the reaction must contain the struck quark and the
unfavoured fragmentation function should therefore vanish.  This
behaviour is seen in Fig.~\ref{fig:d2-d1} which is a plot of
$D_2(z)/D_1(z)$ from HERMES~\cite{ge98} and EMC~\cite{ar89}.  The data
indicate that $D_2(z)/D_1(z)$ does not drop off with $z$ as
quickly as expected from Eqn.~\ref{eqn:indFF}, especially for
$z>0.7$.  However, the error bars are large and the high~$z$ behaviour 
remains to be confirmed.

\subsection{Polarised Valence and Sea Quark Distributions}\label{subsec:DV}

\subsubsection{Valence Quarks:}

The contribution to the nucleon spin from the valence quarks can be
determined from the polarisation dependent yield of charged pions from
two targets~\cite{fr89}.  We will work through this example in some
detail to show how an apparently complicated quantity can be written
simply in terms of four quark distribution functions.

First, write out the numerator of Eqn.~\ref{eqn:semi} for $\pi^+$
production on a H target when the spins of the beam and the target are
opposite:
\begin{equation}
N_{\uparrow\downarrow}^{\pi^+}= \frac{4}{9}~u^+ D_u^{\pi^+} +
\frac{1}{9} d^+ D_d^{\pi^+} + \frac{4}{9} \bar{u}^+ D_{\bar{u}}^{\pi^+} +
\frac{1}{9} \bar{d}^+ D_{\bar{d}}^{\pi^+} + \frac{1}{9} s^+ D_s^{\pi^+} +
\frac{1}{9} \bar{s}^+ D_{\bar{s}}^{\pi^+}.
\label{eqn:nupdown1}
\end{equation}

\noindent
Using Eqns.~\ref{eqn:favFF}, \ref{eqn:unfavFF}, and \ref{eqn:strangFF},
we can re-write this as:
\begin{equation}
N_{\uparrow\downarrow}^{\pi^+}= \frac{4}{9} u^+ D_1 +
\frac{1}{9} d^+ D_2 + \frac{4}{9} \bar{u}^+ D_2 +
\frac{1}{9} \bar{d}^+ D_1 + \frac{1}{9} s^+ D_3 +
\frac{1}{9} \bar{s}^+ D_3.
\label{eqn:nupdown2}
\end{equation}

\noindent
A similar equation holds for the production of $\pi^-$ on H with the
same relative spin orientations:
\begin{equation}
N_{\uparrow\downarrow}^{\pi^-}= \frac{4}{9} u^+ D_2 +
\frac{1}{9} d^+ D_1 + \frac{4}{9} \bar{u}^+ D_1 + 
\frac{1}{9} \bar{d}^+ D_2 + \frac{1}{9} s^+ D_3 +
\frac{1}{9} \bar{s}^+ D_3.
\label{eqn:nupdown3}
\end{equation}

\noindent
Taking the difference of Eqns.~\ref{eqn:nupdown2} and \ref{eqn:nupdown3},
the strange quark contributions drop out and the term ($D_1 - D_2$) factors
out:
\begin{equation}
N_{\uparrow\downarrow}^{\pi^+} - N_{\uparrow\downarrow}^{\pi^-}= 
\frac{4}{9} (u^+ - \bar{u}^+) (D_1 - D_2)~-~
\frac{1}{9} (d^+ - \bar{d}^+) (D_1 - D_2).
\label{eqn:nupdown4}
\end{equation}

\noindent
Since $(q^+ - \bar{q}^+)$ is just the valence distribution for quark
flavour q, Eqn.~\ref{eqn:nupdown4} can finally be written as:
\begin{equation}
N_{\uparrow\downarrow}^{\pi^+} - N_{\uparrow\downarrow}^{\pi^-}= 
\frac{1}{9} (4 u_v^+ - d_v^+) (D_1 - D_2).
\label{eqn:nupdown5}
\end{equation}

\noindent
The sea quark contribution has been cancelled out by taking the
difference of $\pi^+$ and $\pi^-$ production and we are left with an
expression for the valence quarks alone.  A similar expression holds
for the other spin direction, which is sensitive to $u_v^-$ and
$d_v^-$:
\begin{equation}
N_{\uparrow\uparrow}^{\pi^+} - N_{\uparrow\uparrow}^{\pi^-}= 
\frac{1}{9} (4 u_v^- - d_v^-) (D_1 - D_2).
\label{eqn:nupup}
\end{equation}

\noindent
Taking the difference (sum) of Eqns.~\ref{eqn:nupdown5} and 
\ref{eqn:nupup}, we get an expression which depends on the polarised
(unpolarised) valence quark distribution functions:
\begin{equation}
(N_{\uparrow\downarrow}^{\pi^+} - N_{\uparrow\downarrow}^{\pi^-}) -
(N_{\uparrow\uparrow}^{\pi^+} - N_{\uparrow\uparrow}^{\pi^-}) = 
\frac{1}{9} (4 \Delta u_v - \Delta d_v) (D_1 - D_2),
\label{eqn:pidif}
\end{equation}
\begin{equation}
(N_{\uparrow\downarrow}^{\pi^+} - N_{\uparrow\downarrow}^{\pi^-}) +
(N_{\uparrow\uparrow}^{\pi^+} - N_{\uparrow\uparrow}^{\pi^-}) = 
\frac{1}{9} (4 u_v - d_v) (D_1 - D_2).
\label{eqn:pisum}
\end{equation}

\noindent
Finally, the fragmentation functions cancel in the ratio of
Eqns.~\ref{eqn:pidif} and \ref{eqn:pisum} and we get:
\begin{equation}
\frac{(N_{\uparrow\downarrow}^{\pi^+} - N_{\uparrow\downarrow}^{\pi^-}) -
(N_{\uparrow\uparrow}^{\pi^+} - N_{\uparrow\uparrow}^{\pi^-})}
{(N_{\uparrow\downarrow}^{\pi^+} - N_{\uparrow\downarrow}^{\pi^-}) +
(N_{\uparrow\uparrow}^{\pi^+} - N_{\uparrow\uparrow}^{\pi^-})} [{\rm proton}]=
\frac{4 \Delta u_v - \Delta d_v}{4 u_v - d_v}.
\label{eqn:piratioh}
\end{equation}

\noindent
The unpolarised quark distribution functions in the denominator have
been determined from unpolarised DIS experiments so that we
are left with only two unknowns in Eqn.~\ref{eqn:piratioh}.  A similar
expression can be written for a deuterium target:
\begin{equation}
\frac{(N_{\uparrow\downarrow}^{\pi^+} - N_{\uparrow\downarrow}^{\pi^-}) -
(N_{\uparrow\uparrow}^{\pi^+} - N_{\uparrow\uparrow}^{\pi^-})}
{(N_{\uparrow\downarrow}^{\pi^+} - N_{\uparrow\downarrow}^{\pi^-}) +
(N_{\uparrow\uparrow}^{\pi^+} - N_{\uparrow\uparrow}^{\pi^-})} [{\rm deuteron}]=
\frac{\Delta u_v + \Delta d_v}{u_v + d_v},
\label{eqn:piratiod}
\end{equation}

\noindent and we can now solve for $\Delta u_v$ and $\Delta d_v$
separately.  This is a powerful tool to isolate the contribution to
the nucleon spin from the valence quarks.  Note that we must be able
to identify pions for this technique to work.  However, the same
formalism can be applied to all hadrons at the expense of extra model
dependence if Monte Carlo techniques are used to calculate corrections
for the fact that the type of final state hadron is not determined.
Note also that data on {\em two} targets are needed to determine the
polarised valence quark distributions using this technique.

\subsubsection{Sea Quarks:}

Different combinations of identified hadron yields can be used to
isolate different quark distributions.  In particular since $\pi^-$ is
a $\bar{u}d$ state and that the coupling in DIS scattering is four
times larger for the $\bar{u}$ quark than for the $d$ quark,
polarisation dependent $\pi^-$ production should be sensitive to the
polarised $\bar{u}$ distribution.  Furthermore, since $K^-$ is a
$\bar{u}s$ state, it is {\em all sea} and should have increased
sensitivity, especially to the strange sea~\cite{cl91}.  Expressions 
for $\pi^-$ and $K^-$ production based on Eqn.~\ref{eqn:semi} can be 
derived in a similar way to the previous section.

Once high-statistics data sets exist on the production of $\pi^\pm$
and $K^\pm$ on more than one target, a global fitting procedure can be
used to determine all polarised quark distributions at once (see next
section).  It is hoped that such data sets will be available in the
next two years.

\begin{figure}
\hspace{0.75cm}\psfig{figure=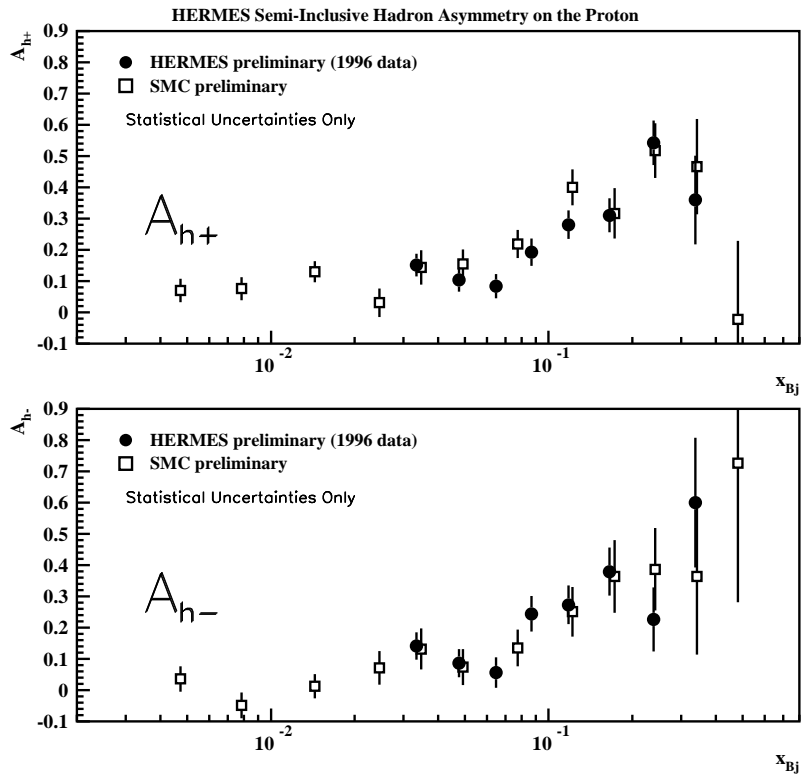,width=4.0in}
\caption{Single hadron asymmetries from SMC$^{52}$ and HERMES$^{53}$
for a H target.
\label{fig:semi-had}}
\end{figure}

\subsection{Single Hadron Asymmetries and Polarised Quark Distributions}

\subsubsection{Single Hadron Asymmetries}

Semi-inclusive data are usually presented in the form of single hadron
asymmetries defined as follows:
\begin{equation}
A_h= \frac{N_{\uparrow\downarrow}^h - N_{\uparrow\uparrow}^h}
{N_{\uparrow\downarrow}^h + N_{\uparrow\uparrow}^h}.
\label{eqn:singlehadasym}
\end{equation}

Data on semi-inclusive processes exist only from HERMES and SMC, with
identified pions only available at HERMES.  The acceptance of the SLAC
spectrometers makes detection of a hadron in coincidence essentially
impossible.  SMC has data for both H and D, while HERMES currently
only has data on H and $^3$He (D will be run in 1998-99).  The charged
hadron single spin asymmetries for the proton are shown in
Fig.~\ref{fig:semi-had} for SMC~\cite{ad98} and HERMES~\cite{me98}.
The error bars are comparable.  However only one third of the HERMES
data are included in this plot.  As in the inclusive case, the SMC
data go to lower $x$ because of the higher beam energy.  The data
quality is already quite good and will be improved when HERMES
finishes the analysis of the 1997 data.  The asymmetries are positive
and rather large, especially at $x \ge 0.1$.  

\begin{figure}
\hspace{0.75cm}\psfig{figure=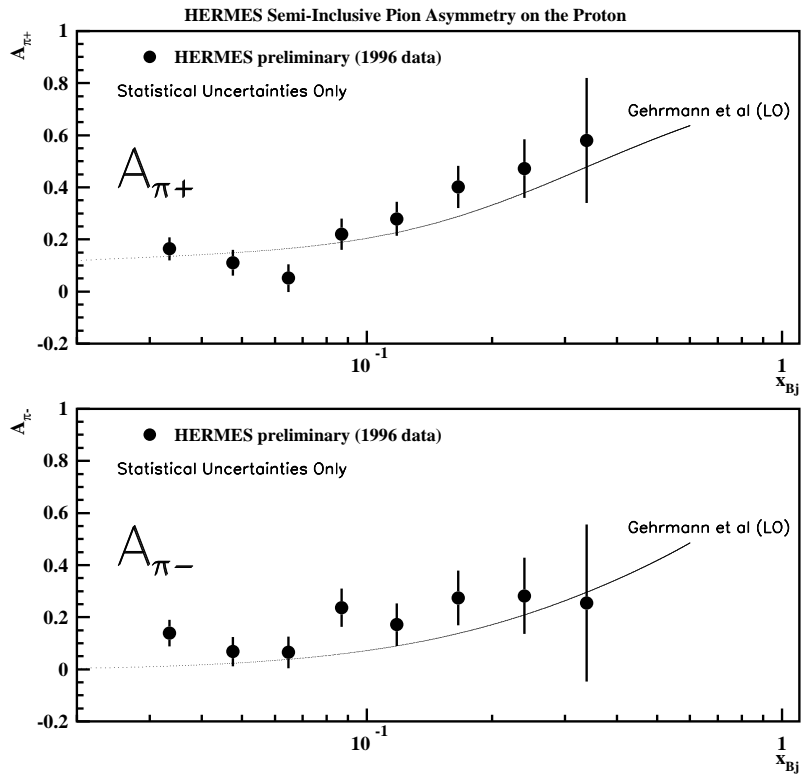,width=4.0in}
\caption{Single pion asymmetries from HERMES$^{53}$ for a H target. The data
are compared to quark distributions from a leading order
fit by Gehrmann and Stirling$^{30}$.
\label{fig:semi-pi}}
\end{figure}

Pion asymmetries from HERMES~\cite{me98} are shown in
Fig.~\ref{fig:semi-pi} where they are compared to the fit of quark
distribution functions by Gehrmann and Stirling~\cite{ge96}.  Note
that here too, only one third of the data have been analysed.

\begin{figure}
\hspace{1.0cm}\psfig{figure=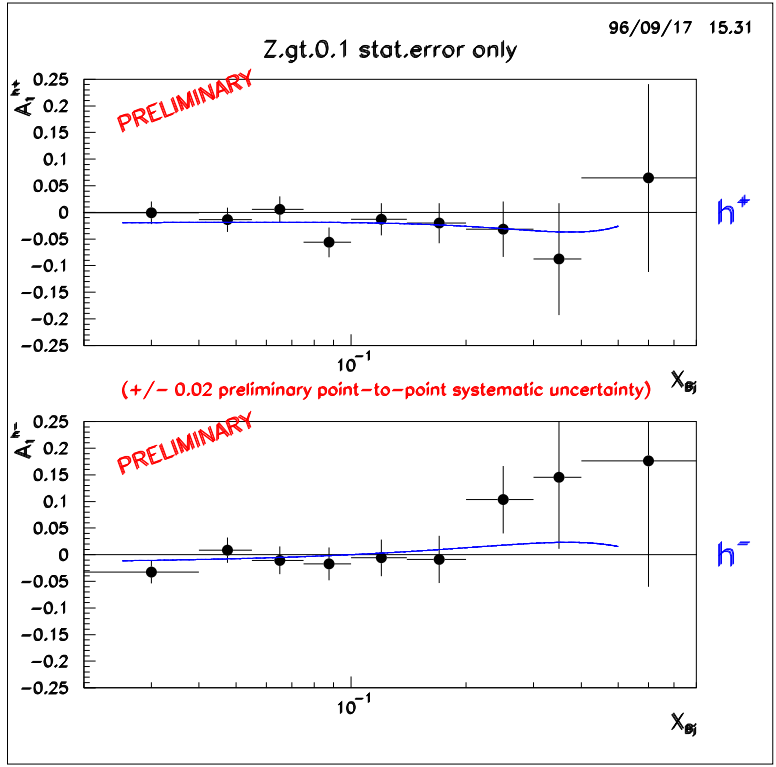,width=4.0in}
\caption{Single hadron asymmetries from HERMES$^{54}$ for a $^3$He target.
\label{fig:semi-he3}}
\end{figure}

Finally, Fig.~\ref{fig:semi-he3} shows the hadron asymmetries measured
by HERMES~\cite{sc97} on a $^3$He target in 1995.  Since this was the
commissioning year for the experiment, the statistics are quite low
and the error bars are correspondingly large.  However, it is clear
that the asymmetries are much smaller for the neutron than for the
proton as in the inclusive case.  The data set for the neutron will be
improved significantly in 1998-99 when HERMES runs with a deuterium
target.

\subsubsection{Polarised Quark Distributions; the Purity Method}

The best way to determine the polarised quark distribution functions
has been to perform a global fit to the inclusive data.  The same can
be done including the semi-inclusive data when a sufficiently large
data base has been accumulated.  We can write the single hadron
asymmetries in terms of quark distribution functions and fragmentation
functions:
\begin{equation}
A_1^h= \frac{1}{D}\, \frac{N_{\uparrow\downarrow}^h - N_{\uparrow\uparrow}^h}
{N_{\uparrow\downarrow}^h + N_{\uparrow\uparrow}^h} =
\frac{\sum_{i} e_i^2\,\ \Delta q_i(x,Q^2)\,\ D_i^h(z,Q^2)}{\sum_{i} e_i^2\,\ 
q_i(x,Q^2)\,\ D_i^h(z,Q^2)}~(1 + R(x,Q^2)),
\label{eqn:hadasym}
\end{equation}

\noindent
where the sum is over quarks and anti-quarks separately and 
$R= \sigma_L / \sigma_T$ as seen earlier.  The $(1 + R)$ term is
needed to be consistent with the extraction of the unpolarised
quark distribution functions using $F_2$ which does not assume $R=0$.
We can define functions $P_i^h$, called {\em Purities}, as the
probability that a hadron $h$ was produced by the fragmentation of a
quark $q_i$ that was struck in the DIS process~\cite{fu98}.  These
are similar to the fragmentation functions but should not be confused
with them.  The purities are written as follows:
\begin{equation}
P_i^h(x,Q^2,z)= \frac{e_i^2\,\ q_i(x,Q^2)\,\ \int_{0.2}^1 D_i^h(z)\,dz}
{\sum_{j} e_j^2\,\ q_j(x,Q^2)\,\ \int_{0.2}^1 D_j^h(z)\,dz}.
\label{eqn:purities}
\end{equation}

\noindent
We can now re-write Eqn.~\ref{eqn:hadasym} in terms of purities and 
quark polarisations:
\begin{equation}
A_1^h(x)= \sum_q P_q^h(x)\,\ \frac{\Delta q(x)}{q(x)}\ (1 + R(x)).
\label{eqn:dqpurity}
\end{equation}

\noindent
The purities can be calculated by Monte Carlo and hence can also
contain information about the acceptance of the spectrometer and
detector effects.  Purities can also be defined for inclusive
asymmetries as follows:
\begin{equation}
P_q^e(x,Q^2)= \frac{e_q^2\,\ q(x,Q^2)}
{\sum_{q} e_q^2\,\ q(x,Q^2)},
\label{eqn:purityinc}
\end{equation}

\noindent
and give the probability that the inclusive scattering came from quark 
$q$.  If one combines inclusive and semi-inclusive measurements, 
Eqn.~\ref{eqn:dqpurity} becomes a matrix equation and we can determine
the polarised quark distribution functions by inverting the purity matrix.
For example, consider the asymmetries for inclusive positrons ($A_1^{e^+}$),
and charged pion production ($A_1^{\pi^+}$ and $A_1^{\pi^-}$):
\begin{equation}
\left( \begin{array}{c} 
A_1^{e^+} \\ A_1^{\pi^+} \\ A_1^{\pi^-} 
\end{array} \right) =
\left( \begin{array}{ccc}
P_u^e       & P_d^e       & P_{\bar{q}}^e \\
P_u^{\pi^+} & P_d^{\pi^+} & P_{\bar{q}}^{\pi^+} \\
P_u^{\pi^-} & P_d^{\pi^-} & P_{\bar{q}}^{\pi^-}
\end{array} \right)
\left( \begin{array}{c}
\Delta u/u \\ \Delta d/d \\ 
\Delta \bar{q}/\bar{q}
\end{array} \right).
\label{eqn:dqmatrix}
\end{equation}

\noindent
The $(1 + R)$ term is now included in the matrix $P$.
More data can be included to get a larger matrix equation and
determine more quark distribution functions.  
\begin{figure}
\hspace{0.25cm}\psfig{figure=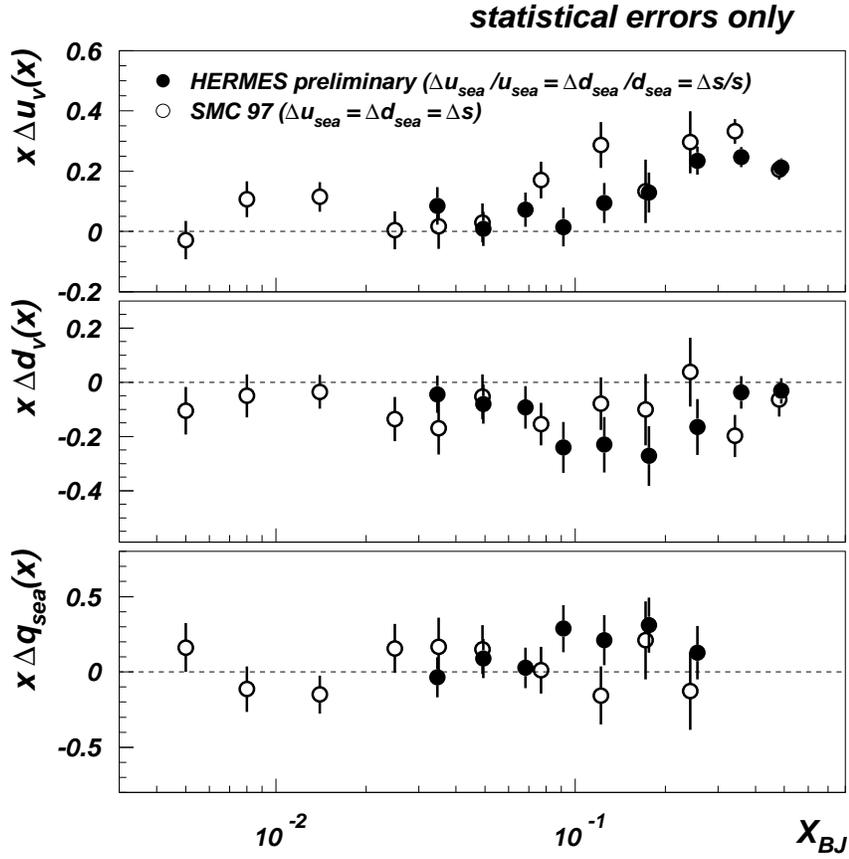,width=4.5in}
\caption{Polarised quark distributions from a fit to SMC data~$^{52}$
and using the purity method for HERMES 1996 data~$^{55}$.
Distributions are plotted for valence $u$ and $d$ quarks as well as
for the sea $\bar{q}$ under the assumption that the polarisation of
the $u_{sea}$, $d_{sea}$, and $s$ quarks is the same.
\label{fig:dq}}
\end{figure}
Care must be taken when inverting the purity matrix because the data
are correlated (e.g.  the semi-inclusive data are a sub-set of the
inclusive data) and the uncertainties must be calculated correctly.
For more details, see the contribution to this workshop by M.-A.~Funk.

Results from SMC~\cite{ad98} and HERMES (1996 only)~\cite{fu98} are
shown in Fig.~\ref{fig:dq} for the polarised valence quark
distributions as well as for the sea distribution $\bar{q}$.  It is
not possible with the current data to separate the contribution from
each type of sea quark.  New data to be taken in the next two years at
HERMES with the RICH detector and with two targets (D and H) will
greatly improve the situation.

\subsection{Charm Production and the Gluon Contribution}\label{subsec:DG}

\begin{figure}
\hspace{3.0cm}\psfig{figure=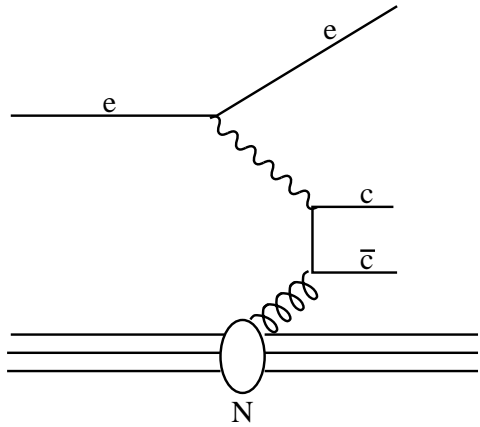,width=2.5in}
\caption{Diagram of the photon-gluon fusion process.  The photon and
the gluon couple through the intermediary of a $c\bar{c}$-pair.
\label{fig:pgfusion}}
\end{figure}

While data collected to date on inclusive and semi-inclusive polarised
DIS are sensitive to the spins of the quarks, we saw earlier that it
is now crucial to determine if the gluons are polarised.  In order to
be sensitive to gluon polarisation, the normal DIS diagrams must be
suppressed.  One way to do this is by considering charm production.
This has proven to be useful in the unpolarised case~\cite{al91}.
There should not be a large charm component in the nucleon
wavefunction (intrinsic charm) so it is not very probable to strike a
charm quark directly.  Furthermore, charm production in the
fragmentation of light quarks is suppressed by the mass of the charm
quark.  As a result, charm production in DIS is dominated by photon-gluon
fusion which is shown diagrammatically in Fig.~\ref{fig:pgfusion}.  In
this process, the virtual photon and the gluon interact through the
intermediary of a $c\bar{c}$ pair.  The charm quarks can manifest
themselves as charmed mesons in the final state (open charm
production: D-mesons), or as J/$\psi$ mesons (hidden charm).  In the
latter case, a second gluon must be emitted by the $c\bar{c}$ system
in order to get a colour singlet final state.  If this gluon is soft,
the interpretation of the process becomes more difficult.  Therefore,
conditions must be placed on the kinematics of the reaction to ensure
that we have inelastic J/$\psi$'s in the final state. The asymmetry in
charm production for the two relative spin states of the beam and
the target can be written as:
\begin{equation}
A_{LL}(z,p_T)= \hat{a}_{LL} \frac{\Delta g}{g}
\end{equation}

\begin{figure}
\hspace{2.0cm}\psfig{figure=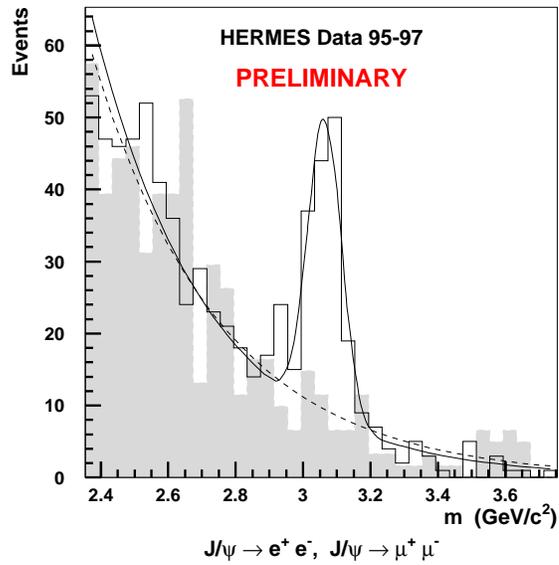,width=3.0in}
\caption{J/$\psi$ mass peak in the e$^+$-e$^-$ and $\mu^+$-$\mu^-$ 
decay channels at HERMES.  The efficiency for $\mu$ identification
will be significantly improved for 1998.
\label{fig:jpsi}}
\end{figure}

\noindent
where $p_T$ is the transverse momentum of the charm meson, and
$\hat{a}_{LL}$ is the asymmetry for the elementary hard scattering
process (i.e. $\gamma-c$).  The asymmetry depends on the production
mechanism.  It has been calculated for J/$\psi$ production in the
colour singlet model~\cite{be81} by Guillet~\cite{gu88} and is
$\approx$ 0.5 for HERMES kinematics.  $\hat{a}_{LL}$ has also been
estimated in the colour octet model which might be more applicable at
high z.  In this case $\hat{a}_{LL}$=1 so that the asymmetry is likely
to be large in either model.  J/$\psi$'s have been clearly identified
at HERMES (Fig.~\ref{fig:jpsi}) and the projected accuracy on $\Delta
g/g$ is 0.52 after the '98-'99 running period and 0.38 by
2001~\cite{am97}.  This will be reduced if improvements to the target
can be made.  While this accuracy is limited, HERMES should be able to
make a qualitative statement on whether $\Delta g$ is non-zero.

An advantage of open-charm production is the larger cross-section
compared to J/$\psi$ production.  This is particularly important at
HERMES where the c.m. energy is barely above threshold.  On the other
hand, the identification of D-mesons is more difficult than
J/$\psi$, especially without $K/\pi$ separation in the hadron PID.

\section{The Future of the Field}\label{sec:future}

HERMES is the only running polarised DIS experiment for the moment.
SLAC has approved an extension to E155 to collect high statistics data
with a transversely polarised target to improve the data on $g_2$.
This experiment should run in 1999.  An experiment called COMPASS is
being prepared at CERN to measure semi-inclusive DIS, in particular
open charm production at higher beam energy than HERMES to study the
gluon polarisation.  The capability to have polarised protons in RHIC
is being included in the accelerator design.  The physics to be
addressed in high energy $\vec{p}$-$\vec{p}$ collisions is described
briefly below and in another contribution to these proceedings~\cite{ha98}.
Finally, DESY is studying the possibility to accelerate polarised
protons in HERA to study polarised DIS at very small values of $x$.

\subsection{HERMES Upgrades}

\begin{figure}
\begin{turn}{90}
\hspace{1.0cm}\psfig{figure=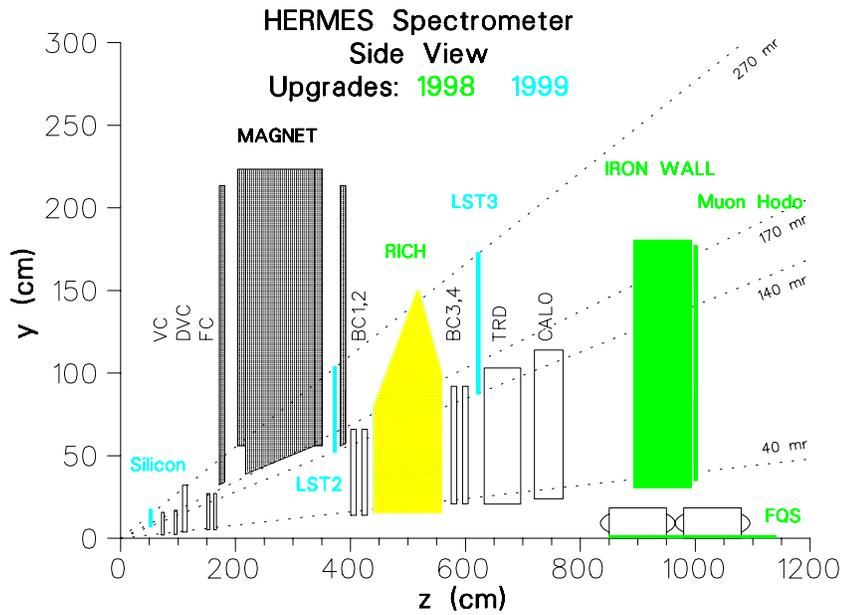,height=11.0cm,width=8.0cm}
\end{turn}
\caption{Schematic representation of the upgrades to the HERMES spectrometer.
The RICH, the iron wall for muon identification, and the FQS to
measure the kinematics of electrons from low $Q^2$ events will all be
ready for the 1998 running period.  The limited streamer tubes to
increase the acceptance for $\mu^+$-$\mu^-$ pairs and most of the
silicon-strip detectors in the target chamber will not be ready until
1999.  
\label{fig:hermes-upgrade}}
\end{figure}

HERMES has embarked on an upgrade program for the 1998 run and beyond
to include kaon identification and improve the efficiency for the
detection of charm mesons~\cite{am97,ci97,he97}.   A schematic
of the HERMES upgrades is shown in Fig.~\ref{fig:hermes-upgrade}.

Hadron identification will be greatly improved by replacing the
threshold \v{C}erenkov detector by a Ring Imaging \v{C}erenkov
detector (RICH)~\cite{ci97}.  The RICH will use a dual radiator
system: $\rm C_4F_{10}$ gas with an index of refraction of 1.0014 and
clear aerogel with n=1.03.  The \v{C}erenkov photons will be detected
by an array of 3/4" phototubes (roughly 2000 per half). The RICH will
provide identification of pions, kaons, and protons (and $\rm \bar{p}$)
over most of the kinematic range of the experiment (2-20~GeV).  This
will allow HERMES to measure all hadron asymmetries and thus determine
the contribution of each quark species to the nucleon's spin.
Furthermore, kaon identification will allow the detection of D-mesons
through the K-$\pi$ decay channel.

Further improvements to the charm detection efficiency will be made on
three fronts~\cite{am97}.  First, an iron wall to absorb hadrons has
been erected behind the calorimeter followed by a scintillator
hodoscope.  This system will improve the muon identification and
double the number of J/$\psi$ particles detected by including the $\mu
^+ \mu ^-$ decay channel.  Better muon identification will also help
for semi-leptonic D-meson decays.  The second upgrade related to charm
is a Forward Quadrupole Spectrometer (FQS) which will be installed in
the first two machine quadrupoles downstream of the experiment.  This
device will detect electrons from low-$Q^2$ events, which are
scattered down the beampipe.  This will allow full kinematic
reconstruction of these events and help separate elastic (or
diffractive) J/$\psi$ production from events which can be interpreted
in terms of $\Delta g$.  Finally, the acceptance for
J/$\psi$~$\rightarrow$~$\mu^+\mu^-$ will be increased by extending the
tracking in front of the magnet and instrumenting the corresponding
region behind the magnet steel which will act as a muon filter.  The
enlarged $\mu^+\mu^-$ acceptance will greatly increase the J/$\psi$
sample since most of these particles are produced with very low energy
and hence with a large opening angle for the decay muons. 

HERMES is also preparing two arrays of silicon-strip detectors to be
located inside the target chamber~\cite{he97}: two wheel-shaped
detectors between the storage cell and the exit window
($\Lambda$-wheels), and recoil detectors parallel to the storage cell.
The recoil detector has many functions.  It can detect slow recoil
protons in e-d scattering to tag e-n events and eliminate nuclear
corrections.  The detection of a recoil proton in J/$\psi$-production
can be used to eliminate elastic events and also to help reconstruct
decays where only one of the leptons is in the spectrometer
acceptance.  As their name implies, the $\Lambda$ wheels will be used
to study $\Lambda$ production and the subsequent decay to
$p$~+~$\pi^-$ ($\Lambda$) or $\bar{p}$~+~$\pi^+$ ($\bar{\Lambda}$).
The idea is to take advantage of the fact that the $\Lambda$-decay is
self-analysing making it possible to measure the polarisation of the
$\Lambda$'s.  This polarisation can be written as:
\begin{equation}
P_{\Lambda} = (P_B D + P_T \frac{\Delta u}{u}) \ \ \frac{\Delta u^{\Lambda}}
{u^{\Lambda}}
\end{equation}
where $P_B$ and $P_T$ are the beam and target polarisations
respectively, $D$ is a depolarisation factor and $\Delta u/u$ and
$\Delta u^{\Lambda}/u^{\Lambda}$ are the polarisations of the
$u$-quark in the proton and the $\Lambda$.  With an unpolarised
target, $\Delta u^{\Lambda}/u^{\Lambda}$ can be determined and the
spin transfer from the $u$-quark in the proton to the $\Lambda$ can be
measured.  If we assume that this quantity does not depend on the
target orientation, we can then determine the transverse polarisation
of $u$-quarks in the proton by measuring $P_{\Lambda}$ with a
transversely polarised target.  While the current acceptance of the HERMES
spectrometer is not well matched to $\Lambda$ detection, the
$\Lambda$-wheels will substantially increase the acceptance and also 
extend the range for negative values of $x_F$.

\subsection{COMPASS at CERN}

The COMPASS experiment at CERN is being designed to study
semi-inclusive processes at a higher energy than HERMES (100~GeV
muons)~\cite{co96}.  The goal is to measure $\Delta g /g$ to an
accuracy of 0.11 after 1.5 years of data taking.  The main advantage
over HERMES is the much higher charm production cross-section due to
the higher beam energy.  The principal channels to be studied are:

\centerline{$D^0 \rightarrow K^- \pi^+~~;~~\bar{D^0} 
\rightarrow K^+ \pi^-$.}

\noindent
A two stage spectrometer with very large acceptance is being built.
Good hadron identification will be provided by two RICH detectors.
The targets will be similar to those used at SMC and E155: NH$_3$
(85\% polarisation) and LiD (50\% polarisation).  COMPASS will also
make the semi-inclusive measurements described in Sec.~\ref{sec:semiinc}.
The experiment is expected to start data taking in 2000.

\subsection{Polarised beams at RHIC}

The capability to study collisions of 250~GeV polarised protons at
RHIC is being prepared~\cite{sr92}.  A more complete description of
this program is given in the contribution to these proceedings by
J.~Harris~\cite{ha98} and the reader is referred to this paper.
Briefly, the polarisation of the sea quarks can be measured using the
Drell-Yan process ($q \bar{q} \rightarrow \mu^+ \mu^-$), while
sensitivity to the gluon polarisation can be obtained by studying the
production of isolated photons (the so-called direct photon
process: $q g \rightarrow \gamma q$), jet production, or J/$\psi$
production.  The direct photon experiment is challenging because of
the large background of $\gamma$'s from $\pi^0$ decays.  While it is
possible to isolate $gg$-processes by using the characteristics of
jets (e.g. the invariant mass of the 2-jet system), the interpretation
is more difficult than for the direct photon process.  RHIC is
scheduled to produce heavy ion collisions in 1999 and it is expected
that some time could be set aside for polarised protons as early as
2000.

\subsection{Polarised Protons at HERA}

In the long term ($\approx$2005), DESY is studying the possibility of
accelerating polarised protons in HERA~\cite{de97,de95} which would
allow the extension of the data set on $g_1$ down to very low $x$.
One of the problems associated with this program is the small
depolarisation factor D due to the kinematics of the scattering.
This means that the polarisation of the virtual photons is small and
very high luminosity is required to compensate.  Furthermore, getting
polarised protons through the accelerator chain at DESY is not trivial
and will involve the use of partial Siberian snakes in PETRA and as
many as four full snakes in HERA.  Nevertheless, information on $g_1$
at low $x$ is crucial to our understanding of how to do the
extrapolation to $x=0$.  HERA would extend the minimum $x$ from about
5.10$^{-3}$ down to a few times 10$^{-5}$.  Furthermore, asymmetries
in the production of jets are expected to be large and will give
information on $\Delta$g.

\subsection{Real photon experiments}

Finally, experiments on charm production using real photons have been
proposed both at SLAC (E156) and at DESY (APOLLON).  However, neither
of these has been approved and they are on hold until further notice.

\section{Summary}

A pedagogical overview of the field of nucleon spin physics has been
presented in these lectures.  Ongoing technological advances in
polarised target and beam production have made possible a series of
experiments over the last 10 years to study how the spin of the
nucleon is related to the spins of its constituents.  There now exists
a large body of data on inclusive polarised deep inelastic scattering
from CERN, SLAC, and DESY.  These data indicate that only a small
fraction of the nucleon's spin comes from the spins of the quarks.
While the Bjorken sum rule has been verified to about 10\% accuracy,
the data on the integrals of $g_1(x)$ for the proton and the neutron
disagree with the predictions of the Ellis-Jaffe sum rules by one to
several sigma.  Further progress in this field will be based on
semi-inclusive measurements where the flavour of the struck quark 
can be tagged by the type of hadron produced in the final state.
This will allow the contributions from valence and sea quarks to be
isolated.  The production of charmed mesons will give information on
the polarisation of the gluons.  The HERMES experiment which will run
with a deuterium target in 1998-99 will produce complete data sets
on semi-inclusive pion and kaon production for the proton and the
neutron.  HERMES will also study charm production and should be able
to make a qualitative statement on whether the gluons are polarised.
A precise determination of $\Delta$g must await the new experiments
being planned at CERN (COMPASS) and at Brookhaven (Spin-RHIC).
The next few years promise to be very exciting in this field and we
might finally find a solution to the ``nucleon spin puzzle''.


\section*{Acknowledgments}
It is a pleasure to thank Eric~Belz, Antje~Br\"ull, Ed~Kinney, and
Manuella~Vincter for their comments on the manuscript.  M.~Vincter
also provided some of the figures.  I also thank Ralf~Kaiser,
Felix~Menden, and Philip~Geiger from whose theses I obtained some of
the figures, and my fellow organizers of the institute for taking most
of the organizational work load which allowed me to concentrate on my
lectures.  Finally, I would like to acknowledge the financial support of
the Natural Sciences and Engineering Research Council of Canada.

\ \

\noindent
* The HERMES internal notes mentioned in the references can be obtained 
from the HERMES web pages: \texttt{http://www-hermes.desy.de/notes/}.

\end{document}